\pdfoutput=1

\documentclass[12pt,letterpaper]{article}

\usepackage[T1]{fontenc}
\usepackage[utf8]{inputenc}
\usepackage{lmodern}
\usepackage{amsmath,amssymb,amsthm}

\usepackage{setspace}
\usepackage{microtype}

\usepackage[letterpaper, margin=1.25in]{geometry}

\usepackage[dvipsnames]{xcolor}
\definecolor{LinkBlue}{RGB}{0,73,135}

\usepackage{bm}
\usepackage{mathtools}
\usepackage{bbm}

\usepackage{enumitem}

\usepackage{longtable,booktabs,array,multirow}

\usepackage{graphicx}
\makeatletter
\def\maxwidth{\ifdim\Gin@nat@width>\linewidth\linewidth\else\Gin@nat@width\fi}
\def\maxheight{\ifdim\Gin@nat@height>\textheight\textheight\else\Gin@nat@height\fi}
\makeatother
\setkeys{Gin}{width=\maxwidth,height=\maxheight,keepaspectratio}

\usepackage[font=small,labelfont=bf,labelsep=period]{caption}
\usepackage{subcaption}

\usepackage{fancyhdr}
\fancypagestyle{plain}{\fancyhf{}\fancyfoot[C]{\thepage}}

\usepackage[round,authoryear]{natbib}
\usepackage{bookmark}
\usepackage{url}
\hypersetup{
  pdftitle={Ill-Conditioned Orthogonal Scores in Double Machine Learning},
  pdfauthor={Gabriel Saco},
  colorlinks=true,
  linkcolor=LinkBlue,
  citecolor=LinkBlue,
  urlcolor=LinkBlue,
}

\theoremstyle{plain}
\newtheorem{theorem}{Theorem}[section]
\newtheorem{lemma}[theorem]{Lemma}

\newtheorem{corollary}[theorem]{Corollary}

\theoremstyle{definition}
\newtheorem{assumption}{Assumption}[section]
\newtheorem{definition}{Definition}[section]

\theoremstyle{remark}
\newtheorem{remark}{Remark}[section]

\newcommand{\E}{\mathbb{E}}
\newcommand{\Var}{\mathrm{Var}}

\newcommand{\Prob}{\mathbb{P}}
\newcommand{\R}{\mathbb{R}}

\newcommand{\pto}{\xrightarrow{p}}

\newcommand{\kappaOof}{\widehat{\kappa}_{\mathrm{oof}}}

\newcommand{\thetahat}{\widehat{\theta}}
\newcommand{\etahat}{\widehat{\eta}}
\newcommand{\mhat}{\widehat{m}}
\newcommand{\ellhat}{\widehat{\ell}}
\newcommand{\Jhat}{\widehat{J}}
\newcommand{\Vhat}{\widehat{V}}
\newcommand{\Uhat}{\widehat{U}}
\newcommand{\esssup}{\mathop{\mathrm{ess\,sup}}}

\newcommand{\maybeincludegraphics}[2][]{\IfFileExists{#2}{\includegraphics[#1]{#2}}{\fbox{Missing: #2}}}

\begin{document}

\title{\textbf{Ill-Conditioned Orthogonal Scores in Double Machine Learning}}

\author{Gabriel Saco\thanks{Universidad del Pac\'ifico. Email: \href{mailto:gsacoalvarado@gmail.com}{gsacoalvarado@gmail.com}. ORCID: \href{https://orcid.org/0009-0009-8751-4154}{0009-0009-8751-4154}}\\[0.5em]
Universidad del Pac\'ifico}

\date{}

\maketitle

\begin{abstract}
\begin{singlespace}
\noindent
Double Machine Learning is often justified by nuisance-rate conditions, yet finite-sample reliability also depends on the conditioning of the orthogonal-score Jacobian. This conditioning is typically assumed rather than tracked. When residualized treatment variance is small, the Jacobian is ill-conditioned and small systematic nuisance errors can be amplified, so nominal confidence intervals may look precise yet systematically under-cover. Our main result is an exact identity for the cross-fitted PLR-DML estimator, with no Taylor approximation. From this identity, we derive a stochastic-order bound that separates oracle noise from a conditioning-amplified nuisance remainder and yields a sufficiency condition for $\sqrt{n}$-inference. We further connect the amplification factor to semiparametric efficiency geometry via the Riesz representer and use a triangular-array framework to characterize regimes as residual treatment variation weakens. These results motivate an out-of-fold diagnostic that summarizes the implied amplification scale. We do not propose universal thresholds. Instead, we recommend reporting the diagnostic alongside cross-learner sensitivity summaries as a fragility assessment, illustrated in simulation and an empirical example.
\end{singlespace}
\end{abstract}

\vspace{0.5em}


\vfill

\thispagestyle{plain}
\newpage

\section{Introduction}
\label{sec:intro}

Double Machine Learning \citep[DML;][]{chernozhukov2018dml} permits the application of machine learning methods for nuisance estimation, while preserving $\sqrt{n}$-inference for low-dimensional treatment effects. This holds because Neyman-orthogonal scores reduce first-order sensitivity to nuisance estimation errors and cross-fitting avoids own-observation bias \citep{chernozhukov2018dml, kennedy2023semipar}. As a result, the corresponding sufficient conditions are typically stated as rate conditions, rather than as parametric smoothness restrictions. In the canonical partially linear regression (PLR) model, for instance, the classical product-rate requirement is $r_n^m r_n^\ell = o_P(n^{-1/2})$, where $r_n^m,r_n^\ell$ denote $L^2$ nuisance-estimation rates \citep{chernozhukov2018dml}.

These conditions control the entry of nuisance estimation error into the orthogonal score. However, these rate conditions do not track how residual treatment variation after conditioning (the PLR analogue of overlap/positivity) directly impacts finite-sample reliability \citep{khanTamer2010overlap,damour2021overlap}. Standard DML theory often relies on regularity conditions to keep this residual variation  bounded away from zero. We instead make the finite-sample sensitivity induced by low residual treatment variation explicit in the canonical PLR-DML setting through an exact identity and a stochastic bound.\footnote{Our main results are for PLR-DML, where the score is affine in $\theta$ and the identity is exact. Appendix~A.1 shows the generic orthogonal-score expansion for nonlinear scores, where a Taylor/linearization step is required.} A simple reportable out-of-fold diagnostic follows as a reporting implication.

An amplification mechanism is transparent in PLR. In this case, the sensitivity of the empirical orthogonal score to the treatment effect $ \theta $ depends on the empirical Jacobian $\widehat{J}_\theta = -\,\widehat{\sigma}_{V}^{2}$, where $V$ is the residualized treatment. If $\widehat{\sigma}_{ V }^{ 2 } $ is small, the score is nearly flat in the $\theta$ direction. Small perturbations of the empirical score equation therefore create large perturbations of $\thetahat$. This corresponds to ill-conditioning in the numerical sense \citep{golubvanloan2013}. We capture this channel through $\kappa := { \sigma }_{ D }^{ 2 }/{\sigma}_{ V }^{ 2 } = 1/(1-R^{ 2 }(D\mid X))$. It scales the impact of any residual bias left after orthogonalization and cross-fitting. While standard errors reflect increased sampling variability when $\sigma_{ V }^{ 2 }$ is small, they do not consider the bias-amplification induced by large $ \kappa $. Coverage can fail as a result, even when reported standard errors are only moderately inflated.

Our first main result is an exact finite-sample decomposition for DML in PLR:
\[
\thetahat-\theta_0 \;=\; \widehat{\kappa}\,(S_n' + B_n').
\]
Here $\widehat{\kappa}$ is the sample analogue of the condition number, $S_n'$ is the standardized oracle
sampling term, and $B_n'$ aggregates the nuisance-driven bias components that remain after orthogonalization
and cross-fitting. The identity is exact. In other words, there is no Taylor approximation, because the PLR score is affine in
$\theta$. This makes the role of $\widehat{\kappa}$ as a finite-sample amplification factor explicit.

Our second main result converts the exact identity into a stochastic-order bound:
\[
\thetahat-\theta_0
= O_P\!\left(\frac{\sqrt{\kappa_n}}{\sqrt{n}} \;+\; \kappa_n\cdot \mathrm{Rem}_n\right),
\qquad
\mathrm{Rem}_n := r_n^m r_n^\ell + (r_n^m)^2 + \frac{r_n^m + r_n^\ell}{\sqrt{n}}.
\]
The first term is the oracle sampling component under conditioning. The second term shows that conditioning
enters multiplicatively with the full nuisance remainder. As a consequence, a sufficient condition for the usual $\sqrt{n}$-approximation (oracle term dominates the remainder) is $\kappa_n\cdot \mathrm{Rem}_n = o(n^{-1/2})$. When overlap is
stable so that $\kappa_n$ is bounded, this reduces to familiar remainder-rate restrictions. If overlap is
weak and $\kappa_n$ is large, the same nuisance errors can be amplified into first-order bias.

Two theoretical connections improve interpretation. First, we link conditioning to semiparametric efficiency
geometry by proving $\kappa = \sigma_D^2 \|\alpha_0\|_{L^2}^2$, where $\alpha_0$ is the Riesz representer
\citep{chernozhukov2022riesz}. Second, we formalize weakening overlap through a triangular-array framework in
which $\kappa_n\to\infty$. This shows that standard $\sqrt{n}$-asymptotics may fail even when nuisance rates are
favorable. The theory also yields an estimable reporting implication. We propose reporting the
condition number,
\[
\kappaOof = \frac{1}{1-\widehat{R}^2_{+}(D\mid X)},
\qquad
\widehat{R}^2_{+} := \max\{0,\widehat{R}^2_{\mathrm{oof}}\},
\]
a scale-invariant measure of residual treatment variation, where
$\widehat{R}^2_{\mathrm{oof}}$ is an out-of-fold predictive $R^2$. When multiple first-stage learners are compared, $\kappaOof$ is computed learner-by-learner.\footnote{Out-of-fold $R^2$ avoids in-sample overfitting bias and aligns the conditioning diagnostic with the cross-fitting used in DML. In PLR, the amplification factor in the bound satisfies $\kappa=\sigma_D^2/\sigma_V^2=1/\{1-R^2(D\mid X)\}$, so $\kappaOof=1/\{1-\widehat R^2_{+}\}$ is a plug-in proxy for the same conditioning object.} This operationalizes the overlap/conditioning component that enters the bound and helps interpret cross-learner sensitivity predicted by the
theory. Importantly, $\kappaOof$ concerns the conditioning of the orthogonalized score, but it is not a test of unconfoundedness and does not by itself validate causal identification. Analogous to weak instruments, where weak first-stage  amplifies bias and motivates strength diagnostics
\citep{staigerstock1997weakiv,stockyogo2005weakiv,moreira2003conditional}, limited residual treatment variation in DML implies large
$\kappa$, which amplifies residual nuisance bias in the orthogonal score equation.

The paper proceeds as follows. Section~\ref{sec:literature} situates our contribution in the literature.
Section~\ref{sec:theory} develops the exact decomposition and stochastic-order bound. Section~\ref{sec:regimes}
characterizes conditioning regimes and rate implications. Section~\ref{sec:simulations} presents Monte Carlo
evidence validating the theoretical predictions. Section~\ref{sec:application} provides an empirical
illustration. Section~\ref{sec:conclusion} concludes.

\section{Related Work}
\label{sec:literature}

\citet{chernozhukov2018dml}
formalize DML as Neyman-orthogonal, cross-fitted score estimation, building on Robinson's partialling-out construction for PLR \citep{robinson1988plr} and influence-function theory
\citep{newey1990semipar}. Subsequent work develops finite-sample guarantees for orthogonal-score estimators
\citep{chernozhukov2023simple}. \citet{kennedy2023semipar} reviews
influence-function--based semiparametric inference (including cross-fitted one-step/TMLE constructions and
doubly robust structure) and \citet{chernozhukov2021rieszregression} develops a geometric debiasing view via
Riesz representers. We build on this literature, but shift attention to a typically implicit PLR regularity.

Standard orthogonal-score DML analyses pair nuisance-rate conditions with a nondegeneracy condition for the score equation. In PLR, this amounts to requiring residualized treatment variation so the score
Jacobian is not close to singular \citep{chernozhukov2018dml}. When this curvature deteriorates, inference becomes a
weak-identification/ill-conditioning problem. Estimating equations flatten and small singular values act as error
amplifiers \citep{kaji2021weakid,hanmccloskey2019singular,breunig2020illposed}. Yet in applied DML this stability
component is rarely tracked or reported alongside the final estimate. We make conditioning an explicit channel in PLR-DML by
isolating a condition-number factor $\kappa$ that governs amplification of the orthogonal-score remainder as residual
treatment variation shrinks. This links overlap-driven fragility in DML to classical collinearity diagnostics 
\citep{belsley1980diagnostic}.

Overlap matters through residualized treatment variation, which governs both identification strength and
finite-sample sensitivity. For binary treatment, efficiency theory implies sensitivity increases as propensity scores
approach 0 or 1 \citep{hahn1998}, motivating strict-overlap conditions and practical remedies such as trimming
\citep{crumpetal2009} and overlap weighting \citep{lietal2018overlap}. When overlap fails, effects may be only partially
identified \citep{khanTamer2010overlap}, and overlap can deteriorate with covariate dimension, creating tension between
flexible adjustment and identification strength \citep{damour2021overlap}. We use this overlap literature primarily to clarify the mechanism by which weak overlap undermines PLR-DML, while leaving the analysis of overlap-adjusted estimators to future work.

Modern causal estimators rely on regularized, data-adaptive learners for nuisance components, so
finite-sample orthogonal-score remainders need not be negligible even with cross-fitting.
High-dimensional and debiased approaches for learned nuisances are developed in
\citet{bellonietal2014highdim}, and sufficient conditions for valid semiparametric inference with
flexible learners are studied in \citet{farrellliangmisra2021}.  Related
robustness ideas are central in doubly robust and targeted learning methods
\citep{kangschafer2007,bangrobins2005,kennedy2023semipar}. These results primarily govern how small nuisance error must be. We complement them by tracking how that error is scaled
by the sample geometry of the PLR score equation. Because this scaling depends on the particular
out-of-fold nuisance fits, our diagnostic is inherently specification-specific for each learner.

\section{Theoretical Results}
\label{sec:theory}

\begin{table}[!tb]
\centering
\caption{Notation and Objects}
\label{tab:notation}
\begin{tabular}{l p{12cm}}
\toprule
\textbf{Symbol} & \textbf{Definition} \\
\midrule
$(Y, D, X)$ & Outcome, treatment, covariates \\
$\theta_0$ & Target causal parameter (constant marginal effect) \\
$g_0(X)$ & Nuisance function in the outcome equation \\
$m_0(X)$ & Treatment regression: $\E[D \mid X]$ \\
$\ell_0(X)$ & Outcome regression: $\E[Y \mid X] = \theta_0 m_0(X) + g_0(X)$ \\
$V$ & Treatment residual: $D - m_0(X)$ \\
$\varepsilon$ & Outcome error: $Y - \theta_0 D - g_0(X)$ \\
$\sigma_V^2, \sigma_D^2$ & $\E[V^2]$ (residual variance), $\Var(D)$ (total treatment variance) \\
\midrule
$\kappa$ & Condition number: $\kappa=\sigma_D^2/\sigma_V^2 = 1/(1-R^2(D\mid X))$ \\
$\kappaOof$ & Out-of-fold diagnostic: $\kappaOof := 1/(1-\widehat{R}^2_{+}(D\mid X))$ \\
$\kappa_n$ & Population condition number along a triangular array sequence \\
$\widehat{\kappa}$ & Sample estimate of $\kappa$ from cross-fitted residuals \\
\midrule
$r_n^m, r_n^\ell$ & $L^2$ convergence rates of nuisance estimators $\widehat{m}, \widehat{\ell}$ \\
$\mathrm{Rem}_n$ & Complete remainder: $r_n^m r_n^\ell + (r_n^m)^2 + (r_n^m + r_n^\ell)/\sqrt{n}$ \\
$S_n, B_n$ & Oracle sampling term, nuisance bias term (Lemma~\ref{lem:bias_decomp}) \\
$S_n', B_n'$ &
Standardized versions: $S_n' := S_n/\widehat{\sigma}_D^2$ and $B_n' := B_n/\widehat{\sigma}_D^2$ \\
$\alpha_0$ & Riesz representer: $V/\sigma_V^2$ (Theorem~\ref{thm:riesz_kappa}) \\
\bottomrule
\end{tabular}
\end{table}

This section introduces notation and assumptions for PLR--DML and develops our finite-sample characterization of how
limited residual treatment variation affects inference.\footnote{We keep the main bound in stochastic-order form because fully nonasymptotic Gaussian approximation requires additional constants and concentration tools, such as finite-sample analyses for sample-splitting/DML-type estimators.} We present (i) an explicit link between overlap
strength and a condition number $\kappa$, (ii) an exact finite-sample identity for $\thetahat-\theta_0$, and (iii) a
stochastic-order bound that yields an operational sufficiency condition for $\sqrt{n}$-valid inference. We write $\kappa$ for the population condition number under a fixed law $P$, and $\kappa_n$ under a triangular-array sequence $(P_n)_{n\ge1}$.

\subsection{Data Structure}

\begin{assumption}[Data Structure]\label{ass:data}
For each $n$, we observe $\{W_{i,n}\}_{i=1}^n$ i.i.d.\ under $P_n$, where $W_{i,n} = (Y_{i,n}, D_{i,n}, X_{i,n})$ consists of outcome $Y \in \R$, treatment $D \in \R$, and covariates $X \in \R^p$. The classical i.i.d.\ setting is $P_n \equiv P$. We suppress index $n$ when unambiguous.
\end{assumption}

\begin{definition}[Function Spaces]\label{def:norms}
For measurable $f: \mathcal{X} \to \R$, define $\|f\|_{L^2} := (\E[|f(X)|^2])^{1/2}$ and $\|f\|_n := (n^{-1}\sum_{i=1}^n f(X_i)^2)^{1/2}$. We employ standard stochastic order notation: $Z_n = O_P(a_n)$ if $|Z_n/a_n|$ is bounded in probability; $Z_n=o_P(a_n)$ if $Z_n/a_n \pto 0$.
\end{definition}

\textbf{Assumption map.} Lemma~\ref{lem:id} uses only $\E[V\varepsilon] = 0$ and $\sigma_V^2 > 0$.
The exact decomposition (Theorem~\ref{thm:exact_decomp}) is algebraic under PLR and does not require rate conditions.
Finite-sample bounds (Theorem~\ref{thm:fs_bound}) additionally use moment bounds, nuisance $L^2$ rates, and residual-variance stability.
Regime analysis allows $\sigma_{V,n}^2 \to 0$ via the triangular array setup.

\subsection{The Partially Linear Regression Model}

The PLR model \citep{robinson1988plr} specifies:
\begin{align}
Y &= \theta_0 D + g_0(X) + \varepsilon, \label{eq:plr_y}\\
D &= m_0(X) + V, \label{eq:plr_d}
\end{align}
where $\theta_0 \in \R$ is the target parameter, $g_0: \mathcal{X} \to \R$ is a nuisance function, $m_0(X) := \E[D \mid X]$ is the treatment regression (conditional mean), and $V := D - m_0(X)$ is the treatment residual satisfying $\E[V \mid X] = 0$ by construction.

\begin{remark}[Terminology]\label{rem:terminology}
When $D$ is binary, $m_0(X) = \E[D \mid X]$ equals the propensity score $e(X)$ \citep{rosenbaumrubin1983}. When $D$ is continuous, $m_0(X)$ is the treatment regression (conditional mean), and our overlap notion is expressed through $\Var(D \mid X)$.
\end{remark}

\subsection{Target Parameter and Identification}

\begin{definition}[Target Parameter]\label{def:target}
The target parameter is the constant marginal effect $\theta_0$ in the partially linear model $Y = \theta_0 D + g_0(X) + \varepsilon$.
\end{definition}

\begin{lemma}[Identification of $\theta_0$]\label{lem:id}
Under the PLR model \eqref{eq:plr_y}--\eqref{eq:plr_d} and Assumption~\ref{ass:causal_plr}, the target parameter $\theta_0$ is identified as the unique solution to
\[
\E\big[V\{Y - \theta D - g_0(X)\}\big] = 0,
\]
equivalently,
\[
\theta_0 = \frac{\E[VY]}{\E[V^2]} = \frac{\E[(D - m_0(X))Y]}{\E[(D - m_0(X))^2]},
\]
provided $\sigma_V^2 = \E[V^2] > 0$. This ``partialling-out'' identification strategy dates to \citet{frischwaugh1933} and \citet{robinson1988plr}.
\end{lemma}

\begin{proof}
From \eqref{eq:plr_y}, $Y = \theta_0 D + g_0(X) + \varepsilon$. Multiply by $V = D - m_0(X)$ and take expectations:
\[
\E[VY] = \theta_0 \E[VD] + \E[Vg_0(X)] + \E[V\varepsilon].
\]
Now $\E[VD] = \E[V(m_0(X) + V)] = \E[V^2] = \sigma_V^2$, and $\E[Vg_0(X)] = \E[\E[V \mid X]g_0(X)] = 0$ since $\E[V \mid X] = 0$. Finally, Assumption~\ref{ass:causal_plr} implies $\E[V\varepsilon] = 0$. Rearranging yields the formula.
\end{proof}

\subsection{Causal Setup and Identification}

\begin{assumption}[Causal PLR and Conditional Mean Independence]\label{ass:causal_plr}
There exist potential outcomes $\{Y(d): d \in \mathcal{D}\}$ \citep{rubin1974,rubin2005} such that
\[
Y(d) = \theta_0 d + g_0(X) + \varepsilon, \qquad \E[\varepsilon \mid D, X] = 0,
\]
and consistency holds: $Y = Y(D)$.
\end{assumption}

\begin{remark}[Causal interpretation and orthogonality]\label{rem:what_identified}
Under Assumption~\ref{ass:causal_plr}, $\theta_0$ is a constant causal marginal effect (a constant CATE). All results below rely on the residual orthogonality moment $\E[V\varepsilon]=0$, which is implied by the causal restriction $\E[\varepsilon\mid D,X]=0$ because
$\E[V\varepsilon]=\E\{\E[(D-\E[D\mid X])\varepsilon\mid X]\}
=\E\{\E[D\varepsilon\mid X]-\E[D\mid X]\E[\varepsilon\mid X]\}=0$.

\end{remark}

\textbf{Causal interpretation.} The estimand $\theta_0$ is causal under standard potential-outcome conditions: consistency ($Y = Y(D)$), conditional exogeneity ($\E[\varepsilon \mid D, X] = 0$), and overlap ($\Var(D \mid X) > 0$). In the PLR framework, these conditions imply the orthogonal moment $\E[V \cdot (Y - g_0(X) - \theta_0(D - m_0(X)))] = 0$ \citep{chernozhukov2018dml}. Our analysis takes this causal target as given and studies how finite-sample inference behaves as overlap weakens (via $\sigma_V$) and $\kappa$ grows. Thus, $\kappa$ measures inferential fragility, not identification validity.

We distinguish (i) strong overlap (fixed-$\kappa$) and (ii) weakening overlap (triangular array):

\begin{assumption}[Strong Overlap]\label{ass:overlap}
There exists $\underline{\sigma}^2 > 0$ such that $\Var(D \mid X = x) \geq \underline{\sigma}^2$ for $P_X$-almost all $x$. This is the continuous-treatment analogue of the positivity condition \citep{rosenbaumrubin1983}. \citet{damour2021overlap} establish that such strict overlap assumptions become more restrictive as covariate dimension grows.
\end{assumption}

\begin{assumption}[Bounded Treatment Variance]\label{ass:bounded_D}
$\sigma_{D,n}^2 \asymp 1$; i.e., $0 < c \leq \sigma_{D,n}^2 \leq C < \infty$ for some constants $c, C$ and all $n$.
\end{assumption}

\begin{remark}[Bounded $\kappa$ under Strong Overlap]\label{rem:bounded_kappa}
Assumption~\ref{ass:overlap} implies $\sigma_{V,n}^2 \geq \underline{\sigma}^2 > 0$. Under bounded treatment variance (Assumption~\ref{ass:bounded_D}), it follows that $\kappa_n \leq (\sup_n \sigma_{D,n}^2)/\underline{\sigma}^2 < \infty$, so $\kappa_n = O(1)$.
\end{remark}

\begin{remark}[Relation to Positivity / Overlap in the Binary Case]\label{rem:positivity}
If $D \in \{0,1\}$ and $e(X) := \Prob(D = 1 \mid X)$, then
\[
\Var(D \mid X) = e(X)\{1 - e(X)\}, \qquad \sigma_V^2 = \E[e(X)\{1 - e(X)\}].
\]
Thus, the usual positivity condition $\epsilon \leq e(X) \leq 1 - \epsilon$ implies $\Var(D \mid X) \geq \epsilon(1 - \epsilon)$, which is a special case of Assumption~\ref{ass:overlap}.
\end{remark}

To analyze regimes where $\kappa$ may grow, we introduce a triangular array framework:

\begin{assumption}[Triangular Array with Weakening Overlap]\label{ass:triangular}
Consider a sequence of DGPs indexed by $n$. There exists a sequence $\underline{\sigma}_n^2 \downarrow 0$ such that for each $n$:
\[
\Var(D_n \mid X_n = x) \geq \underline{\sigma}_n^2 \quad \text{for } P_{X,n}\text{-almost all } x.
\]
Define $\sigma_{V,n}^2 := \E[\Var(D_n \mid X_n)]$ and $\kappa_n := \sigma_{D,n}^2/\sigma_{V,n}^2$.
\end{assumption}

\begin{remark}[Reconciling Fixed and Growing $\kappa$]\label{rem:reconcile}
Assumption~\ref{ass:overlap} covers the fixed-$\kappa$ case (bounded condition number). Assumption~\ref{ass:triangular} covers the growing-$\kappa$ case: as $\underline{\sigma}_n^2 \to 0$, we may have $\sigma_{V,n}^2 \to 0$ and thus $\kappa_n \to \infty$. The rate $\kappa_n = O(n^\gamma)$ for $\gamma \geq 0$ determines the conditioning regime. When presenting results under Assumption~\ref{ass:overlap}, $\kappa$ is treated as fixed; when presenting asymptotic regime analysis, Assumption~\ref{ass:triangular} applies.
\end{remark}

Define $\ell_0(X) := \E[Y \mid X]$ and outcome residual $U := Y - \ell_0(X)$. Under Assumption~\ref{ass:causal_plr}, $\ell_0(X) = \theta_0 m_0(X) + g_0(X)$.

\begin{lemma}[Residual Decomposition]\label{lem:residual}
Under PLR, $U = \theta_0 V + \varepsilon$.
\end{lemma}

\begin{proof}
By definition, $U = Y - \ell_0(X)$. Substituting $Y = \theta_0 D + g_0(X) + \varepsilon$ and $\ell_0(X) = \theta_0 m_0(X) + g_0(X)$:
\begin{align*}
U &= [\theta_0 D + g_0(X) + \varepsilon] - [\theta_0 m_0(X) + g_0(X)] \\
&= \theta_0 D - \theta_0 m_0(X) + \varepsilon \\
&= \theta_0(D - m_0(X)) + \varepsilon = \theta_0 V + \varepsilon. \qedhere
\end{align*}
\end{proof}

\begin{remark}[Interpretation]\label{rem:residual_interp}
Lemma~\ref{lem:residual} shows the outcome residual equals the causal effect times the treatment residual plus noise. The treatment residual $V$ contains all identifying variation. The precision of identifying $\theta_0$ depends on $\Var(V) = \sigma_V^2$.
\end{remark}

\subsection{Variance Components and the Condition Number}

\begin{assumption}[Second Moments]\label{ass:second_mom}
$\E[D^2] < \infty$ and $\E[Y^2] < \infty$ for all $n$.
\end{assumption}

Define the variance components:
\begin{align}
\sigma_D^2 &:= \Var(D), \label{eq:sigmaD}\\
\sigma_V^2 &:= \E[V^2] = \E[\Var(D \mid X)], \label{eq:sigmaV}\\
\sigma_m^2 &:= \Var(m_0(X)). \label{eq:sigmam}
\end{align}

\begin{lemma}[Law of Total Variance]\label{lem:var_decomp}
The law of total variance yields $\sigma_D^2 = \sigma_V^2 + \sigma_m^2$.
\end{lemma}

\begin{proof}
By the law of total variance:
\begin{align*}
\Var(D) &= \E[\Var(D \mid X)] + \Var(\E[D \mid X]) \\
&= \E[(D - m_0(X))^2] + \Var(m_0(X)) \\
&= \sigma_V^2 + \sigma_m^2. \qedhere
\end{align*}
\end{proof}

The population $R^2$ for treatment explained by covariates is:
\begin{equation}\label{eq:R2}
R^2(D \mid X) := \frac{\sigma_m^2}{\sigma_D^2} = 1 - \frac{\sigma_V^2}{\sigma_D^2}.
\end{equation}

\begin{definition}[Condition Number and $R^2$ Representation]\label{def:kappa}
The condition number is:
\begin{equation}\label{eq:kappa}
\kappa := \frac{\sigma_D^2}{\sigma_V^2} = \frac{1}{1 - R^2(D\mid X)}.
\end{equation}

The empirical analogue uses cross-fitted residuals and defines the sample ratio:
\[
\widehat{\sigma}_V^2 := \frac{1}{n}\sum_{i=1}^n \Vhat_i^2, \quad
\widehat{\sigma}_D^2 := \frac{1}{n}\sum_{i=1}^n (D_i - \bar{D})^2, \quad
\bar{D} := \frac{1}{n}\sum_{i=1}^n D_i,
\]
and $\widehat{\kappa} := \widehat{\sigma}_D^2/\widehat{\sigma}_V^2$.
\end{definition}

\begin{remark}[VIF Connection]\label{rem:vif}
The condition number $\kappa = 1/(1 - R^2(D \mid X))$ has the same functional form as the classical Variance Inflation Factor \citep{belsley1980diagnostic}, and reduces to the classical VIF when $R^2(D \mid X)$ is interpreted as the $R^2$ from the linear projection of $D$ on $X$. Our $R^2(D \mid X)$ uses the nonparametric conditional mean $m_0(X) = \E[D \mid X]$, making $\kappa$ a nonparametric generalization.
\end{remark}

\begin{remark}[Properties of $\kappa$]\label{rem:kappa_properties}
By Lemma~\ref{lem:var_decomp}, $\sigma_D^2 \geq \sigma_V^2$, so $\kappa \geq 1$ with equality when $m_0(X)$ is constant (treatment is unpredictable). As $R^2(D \mid X) \to 1$, we have $\sigma_V^2 \to 0$ and $\kappa \to \infty$.
\end{remark}

\begin{remark}[Binary-Treatment Specialization of $\kappa$]\label{rem:kappa_binary}
If $D \in \{0,1\}$ with $\Prob(D = 1) = p$, then $\sigma_D^2 = p(1-p)$ and $\sigma_V^2 = \E[e(X)\{1 - e(X)\}]$, so
\[
\kappa = \frac{p(1-p)}{\E[e(X)\{1 - e(X)\}]},
\]
which grows as overlap weakens.
\end{remark}

\subsection{Cross-Fitting and the DML Estimator}

\begin{definition}[Cross-Fitting]\label{def:crossfit}
A $K$-fold partition $\{I_1, \ldots, I_K\}$ of $\{1,\ldots,n\}$ with disjoint folds. For $i \in I_k$, nuisance estimates $\mhat^{(-k)}, \ellhat^{(-k)}$ are trained on $\{W_j : j \notin I_k\}$. We take $K$ fixed as $n \to \infty$. This sample-splitting strategy originates in \citet{schick1986} and is central to modern semiparametric estimation \citep{chernozhukov2018dml}. Throughout, we assume the relevant second moments exist under $P_n$ so that $\sigma_{D,n}^2$, $\sigma_{V,n}^2$, and $L^2$ norms are well-defined.
\end{definition}

Cross-fitted residuals: $\Vhat_i := D_i - \mhat^{(-k)}(X_i)$, $\Uhat_i := Y_i - \ellhat^{(-k)}(X_i)$ for $i \in I_k$. Define errors:
\begin{align}
\Delta_i^m &:= m_0(X_i) - \mhat^{(-k)}(X_i), \label{eq:delta_m}\\
\Delta_i^\ell &:= \ell_0(X_i) - \ellhat^{(-k)}(X_i). \label{eq:delta_ell}
\end{align}

\begin{remark}[Residual Decomposition]\label{rem:residual_decomp}
With these sign conventions, the cross-fitted residuals decompose as:
\[
\Vhat_i = D_i - \mhat^{(-k)}(X_i) = (D_i - m_0(X_i)) + (m_0(X_i) - \mhat^{(-k)}(X_i)) = V_i + \Delta_i^m,
\]
and similarly $\Uhat_i = U_i + \Delta_i^\ell$. This decomposition is central. Estimated residuals equal true residuals plus nuisance error.
\end{remark}

The PLR score is:
\begin{equation}\label{eq:score}
\psi(W; \theta, \eta) := (D - m(X))\{Y - \ell(X) - \theta(D - m(X))\},
\end{equation}
where $\eta = (\ell, m)$. At true values, $\psi(W; \theta_0, \eta_0) = V\varepsilon$.

\begin{lemma}[Neyman Orthogonality]\label{lem:orthog}
The pathwise derivative of $\E[\psi(W; \theta_0, \eta)]$ with respect to $\eta$ vanishes at $\eta_0$. This property is the cornerstone of debiased machine learning \citep{chernozhukov2018dml}.
\end{lemma}

The proof is in Appendix~A.2. Neyman orthogonality implies the first-order effect of nuisance perturbations on the population moment vanishes. Combined with cross-fitting, the leading sample terms involving $\Delta^m, \Delta^\ell$ are mean-zero and of order $(r_n^m + r_n^\ell)/\sqrt{n}$, while the systematic remainder is second order (e.g., $r_n^m r_n^\ell$ and $(r_n^m)^2$).

\begin{definition}[DML Estimator]\label{def:dml}
\begin{equation}\label{eq:theta_hat}
\thetahat := \frac{\sum_{i=1}^n \Vhat_i \Uhat_i}{\sum_{i=1}^n \Vhat_i^2}.
\end{equation}
\end{definition}

\begin{definition}[Empirical Score Map]\label{def:psi_n}
For any $(\theta, \eta) = (\theta, \ell, m)$ define
\[
\Psi_n(\theta, \eta) := \frac{1}{n}\sum_{i=1}^n (D_i - m(X_i))\{Y_i - \ell(X_i) - \theta(D_i - m(X_i))\}.
\]
With cross-fitting, $\Psi_n(\theta, \etahat)$ is computed using fold-specific $\ellhat^{(-k)}, \mhat^{(-k)}$ for $i \in I_k$.
\end{definition}

\subsection{The Score Jacobian}

\begin{definition}[Empirical Jacobian]\label{def:jacobian}
\begin{equation}\label{eq:jacobian}
\Jhat_\theta := \frac{\partial}{\partial \theta}\left[\frac{1}{n}\sum_i \Vhat_i(\Uhat_i - \theta\Vhat_i)\right] = -\frac{1}{n}\sum_i \Vhat_i^2 = -\widehat{\sigma}_V^2.
\end{equation}
\end{definition}

\begin{remark}[Jacobian Interpretation]\label{rem:jacobian_interp}
The Jacobian magnitude $|\Jhat_\theta| = \widehat{\sigma}_V^2$ measures the score's curvature. In other words, it measures how quickly the score changes as $\theta$ varies. When $\widehat{\sigma}_V^2$ is small, the score is nearly flat in the $\theta$ direction, and small score perturbations cause large parameter shifts. This is the classical numerical-analysis insight that condition numbers govern error propagation \citep{golubvanloan2013}.
\end{remark}

\begin{lemma}[Jacobian-Kappa Relationship]\label{lem:jacobian_kappa}
\begin{equation}
|\Jhat_\theta|^{-1} = \frac{\widehat{\kappa}}{\widehat{\sigma}_D^2}.
\end{equation}
\end{lemma}

\begin{proof}
From Definition~\ref{def:jacobian}, $|\Jhat_\theta| = \widehat{\sigma}_V^2$. From Definition~\ref{def:kappa}, $\widehat{\kappa} = \widehat{\sigma}_D^2/\widehat{\sigma}_V^2$. Therefore:
\[
|\Jhat_\theta|^{-1} = \frac{1}{\widehat{\sigma}_V^2} = \frac{\widehat{\kappa}}{\widehat{\sigma}_D^2}. \qedhere
\]
\end{proof}

\begin{remark}[]\label{rem:key_insight}
When we invert the Jacobian to solve for estimator error, the condition number $\widehat{\kappa}$ appears as a scale-invariant amplification factor. This is the mechanism through which ill-conditioning affects inference.
\end{remark}

\subsection{Connection to the Riesz Representer}

\begin{definition}[Minimal-Norm Representer]\label{def:riesz}
For the PLR moment, we define the minimal-norm representer $\alpha_0: \mathcal{W} \to \R$ as the unique function satisfying:
\begin{enumerate}[label=(\roman*)]
\item $\E[\alpha_0(W) \cdot f(X)] = 0$ for all $f \in L^2(P_X)$;
\item $\E[\alpha_0(W) \cdot D] = 1$;
\item $\|\alpha_0\|_{L^2}^2 = \min\{\|\alpha\|_{L^2}^2 : \alpha \text{ satisfies (i)--(ii)}\}$.
\end{enumerate}
This is the Riesz representer for the functional $\theta \mapsto \E[\alpha(W) \cdot \theta D]$ restricted to the space orthogonal to $L^2(P_X)$, expressed as a constrained minimal-norm problem \citep{chernozhukov2022riesz}.
\end{definition}

\begin{theorem}[Condition Number as Riesz Norm]\label{thm:riesz_kappa}
In the PLR model, $\alpha_0(W) = V/\sigma_V^2$. Also:
\begin{equation}\label{eq:riesz_kappa}
\kappa = \sigma_D^2 \|\alpha_0\|_{L^2}^2,
\qquad \|\alpha_0\|_{L^2}^2 = \sigma_V^{-2}.
\end{equation}
\end{theorem}

\begin{proof}
We show $\alpha_0(W) = V/\sigma_V^2$ satisfies Definition~\ref{def:riesz} and is the unique minimizer.

\textit{Step 1 (Orthogonality):} For any $f \in L^2(P_X)$:
\begin{align*}
\E[\alpha_0(W) \cdot f(X)] &= \frac{1}{\sigma_V^2}\E[V \cdot f(X)] \\
&= \frac{1}{\sigma_V^2}\E[\E[V \mid X] \cdot f(X)] = 0,
\end{align*}
since $\E[V \mid X] = 0$ by construction.

\textit{Step 2 (Normalization):}
\begin{align*}
\E[\alpha_0(W) \cdot D] &= \frac{1}{\sigma_V^2}\E[V \cdot (m_0(X) + V)] \\
&= \frac{1}{\sigma_V^2}(\E[V m_0(X)] + \E[V^2]) = \frac{\sigma_V^2}{\sigma_V^2} = 1.
\end{align*}

\textit{Step 3 (Norm computation):}
\[
\|\alpha_0\|_{L^2}^2 = \E\left[\frac{V^2}{\sigma_V^4}\right] = \frac{\sigma_V^2}{\sigma_V^4} = \frac{1}{\sigma_V^2}.
\]

\textit{Step 4 (Minimality and uniqueness):} Let $\alpha$ be any function satisfying (i)--(ii). Define $h := \alpha - \alpha_0$. Then $h$ satisfies:
\begin{itemize}
\item $\E[h(W) f(X)] = 0$ for all $f \in L^2(P_X)$ (orthogonality inherited);
\item $\E[h(W) D] = 0$ (normalization difference).
\end{itemize}
Since $D = m_0(X) + V$ and $\E[h(W) m_0(X)] = 0$ by orthogonality, we have $\E[h(W) V] = 0$.

Now, $\alpha_0 = V/\sigma_V^2$ is a scalar multiple of $V$, so $\E[h \cdot \alpha_0] = \E[h V]/\sigma_V^2 = 0$. By the Pythagorean identity:
\[
\|\alpha\|_{L^2}^2 = \|\alpha_0 + h\|_{L^2}^2 = \|\alpha_0\|_{L^2}^2 + \|h\|_{L^2}^2 \geq \|\alpha_0\|_{L^2}^2,
\]
with equality if and only if $h = 0$ a.s. Thus $\alpha_0$ is the unique minimizer.

\textit{Step 5 (Final result):}
\[
\sigma_D^2 \cdot \|\alpha_0\|_{L^2}^2 = \sigma_D^2 \cdot \frac{1}{\sigma_V^2} = \kappa. \qedhere
\]
\end{proof}

\begin{remark}[Semiparametric Interpretation]\label{rem:semipar_interp}
Theorem~\ref{thm:riesz_kappa} grounds our diagnostic in modern semiparametric theory \citep{newey1990semipar, kennedy2016semipar,chernozhukov2022riesz}. The minimal-norm representer $\alpha_0$ is the correction weight for double robustness. Its norm measures how much correction is required. A large $\|\alpha_0\|_{L^2}^2$ signals both large variance and large sensitivity to nuisance bias.
\end{remark}

\begin{remark}[Hilbert-Space View]\label{rem:hilbert_view}
Constraint (i) is equivalent to $\E[\alpha(W) \mid X] = 0$, i.e.\ $\alpha$ lies in the orthogonal complement of $L^2(P_X)$ inside $L^2(P)$. The minimization in Definition~\ref{def:riesz} therefore selects the minimum-$L^2$ instrument in the residual variation direction.
\end{remark}

\subsection{The Exact Finite-Sample Decomposition}

Define the oracle sampling term:
\begin{equation}\label{eq:Sn}
S_n := \frac{1}{n}\sum_{i=1}^n V_i \varepsilon_i.
\end{equation}

\begin{lemma}[Bias Decomposition]\label{lem:bias_decomp}
Define the nuisance bias term $B_n := \Psi_n(\theta_0,\etahat)-S_n$. Then
\begin{equation}\label{eq:Bn_expansion}
B_n = B_n^{(1)}+B_n^{(2)}+B_n^{(3)}+B_n^{(4)}+B_n^{(5)},
\end{equation}
with components
\begin{equation}\label{eq:Bn_components}
\begin{aligned}
B_n^{(1)} &:= \frac{1}{n}\sum_i V_i \Delta_i^\ell, \qquad
B_n^{(2)} := -\theta_0 \frac{1}{n}\sum_i V_i \Delta_i^m, \qquad
B_n^{(3)} := \frac{1}{n}\sum_i \Delta_i^m \varepsilon_i, \\
B_n^{(4)} &:= \frac{1}{n}\sum_i \Delta_i^m \Delta_i^\ell, \qquad
B_n^{(5)} := -\theta_0 \frac{1}{n}\sum_i (\Delta_i^m)^2.
\end{aligned}
\end{equation}
\end{lemma}

\begin{proof}
Expand $\Vhat_i(\Uhat_i - \theta_0\Vhat_i)$ using $\Vhat_i = V_i + \Delta_i^m$ and $\Uhat_i = U_i + \Delta_i^\ell = \theta_0 V_i + \varepsilon_i + \Delta_i^\ell$:
\begin{align*}
\Vhat_i(\Uhat_i - \theta_0\Vhat_i) &= (V_i + \Delta_i^m)[(\theta_0 V_i + \varepsilon_i + \Delta_i^\ell) - \theta_0(V_i + \Delta_i^m)] \\
&= (V_i + \Delta_i^m)[\varepsilon_i + \Delta_i^\ell - \theta_0\Delta_i^m].
\end{align*}
Expanding:
\begin{align*}
&= V_i\varepsilon_i + V_i\Delta_i^\ell - \theta_0 V_i\Delta_i^m \\
&\quad + \Delta_i^m\varepsilon_i + \Delta_i^m\Delta_i^\ell - \theta_0(\Delta_i^m)^2.
\end{align*}
Averaging over $i$ and subtracting $S_n = n^{-1}\sum_i V_i\varepsilon_i$ yields the five terms. \qedhere
\end{proof}

\begin{remark}[Interpretation of Bias Components]\label{rem:bias_interp}
Terms $B_n^{(1)}$--$B_n^{(3)}$ are ``first-order'' in nuisance error: under cross-fitting, they have conditional mean zero and contribute $O_P(n^{-1/2})$ variance. Term $B_n^{(4)}$ is the product term driving the product-rate requirement. Term $B_n^{(5)}$ is always negative (for $\theta_0 > 0$) and scales as $(r_n^m)^2$.
\end{remark}

\begin{theorem}[Exact Decomposition]\label{thm:exact_decomp}
Under PLR, the DML estimator satisfies:
\begin{equation}\label{eq:exact_decomp}
\thetahat - \theta_0 = \widehat{\kappa}\,(S_n' + B_n').
\end{equation}
where $S_n' := S_n/\widehat{\sigma}_D^2$ and $B_n' := B_n/\widehat{\sigma}_D^2$. This is an exact algebraic identity. It does not rely on any Taylor expansion or linearization.
\end{theorem}

\begin{proof}
\textit{Step 1 (Solve for estimator):} The DML estimator solves $\Psi_n(\thetahat, \etahat) = 0$. Since the score is affine in $\theta$:
\[
\Psi_n(\theta, \etahat) = \frac{1}{n}\sum_i \Vhat_i\Uhat_i - \theta \cdot \widehat{\sigma}_V^2.
\]
Setting $\Psi_n = 0$: $\thetahat = \frac{1}{n}\sum_i \Vhat_i\Uhat_i / \widehat{\sigma}_V^2$.

\textit{Step 2 (Express error):} Subtracting $\theta_0$:
\[
\thetahat - \theta_0 = \frac{\frac{1}{n}\sum_i \Vhat_i(\Uhat_i - \theta_0\Vhat_i)}{\widehat{\sigma}_V^2} = \frac{\Psi_n(\theta_0, \etahat)}{\widehat{\sigma}_V^2}.
\]

\textit{Step 3 (Decompose score):} By Lemma~\ref{lem:bias_decomp}:
\[
\Psi_n(\theta_0, \etahat) = S_n + B_n.
\]

\textit{Step 4 (Standardize):}
\begin{align*}
\thetahat - \theta_0 &= \frac{S_n + B_n}{\widehat{\sigma}_V^2} \\
&= \frac{\widehat{\sigma}_D^2}{\widehat{\sigma}_V^2} \cdot \frac{S_n + B_n}{\widehat{\sigma}_D^2} \\
&= \widehat{\kappa}(S_n' + B_n'). \qedhere
\end{align*}
\end{proof}

\begin{remark}[Exactness of the decomposition]\label{rem:why_exact}
The decomposition is exact because the PLR score is affine in $\theta$. There is no Taylor expansion, hence no remainder term. This exactness holds at any sample size. Appendix~A.1 records the corresponding generic identity for cross-fitted score estimators where the score need not be affine in $\theta$. 
\end{remark}

\subsection{Variance Inflation versus Bias Amplification}

The exact decomposition shows that $\widehat{\kappa}$ multiplies both the oracle sampling component and the nuisance-induced
bias, but the inferential consequences are different. On the one hand, variance inflation arises
through the oracle term $\widehat{\kappa}\,S_n'$, since $S_n'$ scales with $\sigma_V$ and, under the efficiency bound in
Theorem~\ref{thm:efficiency}, the relevant benchmark variance is $V_{\mathrm{eff}}=\sigma_\varepsilon^2/\sigma_V^2$.
In this case, the usual standard errors track the increased sampling variability induced by limited residual treatment
variation, so coverage can remain approximately nominal. On the other hand, bias amplification arises through
$\widehat{\kappa}\,B_n'$, because any remaining nuisance error is multiplied by $\kappa$. 

\subsection{Stochastic-Order Implications of the Finite-Sample Identity}

\begin{assumption}[Bounded Moments]\label{ass:moments}
$\E[D^4], \E[Y^4] < \infty$.
\end{assumption}

\begin{assumption}[Moment Bounds]\label{ass:moment_bounds}
There exists a constant $C < \infty$ such that:
\begin{enumerate}[label=(\roman*)]
\item $\esssup_x \E[V^2 \mid X = x] \leq C$;
\item $\esssup_{d,x} \E[\varepsilon^2 \mid D = d, X = x] \leq C$.
\end{enumerate}
\end{assumption}

\begin{remark}[Role of Moment Bounds]\label{rem:moment_bounds}
Assumption~\ref{ass:moment_bounds} (i) ensures that $V^2$ has uniformly bounded conditional expectation, enabling the bound $\E[V^2 (\Delta^\ell)^2] \leq C \|\Delta^\ell\|_{L^2}^2$. Assumption~\ref{ass:moment_bounds} (ii) controls the oracle term variance: $\E[V^2 \varepsilon^2] = \E[V^2 \E[\varepsilon^2 \mid D, X]] \leq C \sigma_{V,n}^2$. Conditional homoskedasticity $\E[\varepsilon^2 \mid D, X] = \sigma_\varepsilon^2$ is a special case.
\end{remark}

\begin{assumption}[Nuisance Rates]\label{ass:rates}
$\|\mhat^{(-k)} - m_0\|_{L^2} = O_P(r_n^m)$, $\|\ellhat^{(-k)} - \ell_0\|_{L^2} = O_P(r_n^\ell)$.
Here $r_n^m$ and $r_n^\ell$ are deterministic rate sequences (so $\mathrm{Rem}_n$ is deterministic).
\end{assumption}

\begin{assumption}[Residual-Variance Stability]\label{ass:var_stability}
$\sigma_{V,n}^2 / \widehat{\sigma}_V^2 = O_P(1)$. Equivalently, there exists $c > 0$ with $\Prob(\widehat{\sigma}_V^2 \geq c \sigma_{V,n}^2) \to 1$.
\end{assumption}

\begin{remark}[Role of Variance Stability]\label{rem:var_stability}
Assumption~\ref{ass:var_stability} ensures that the empirical residual variance $\widehat{\sigma}_V^2$ does not collapse relative to the population variance $\sigma_{V,n}^2$. This is needed to control the oracle term scaling $S_n/\widehat{\sigma}_V^2$. Under strong overlap with consistent estimation, this holds automatically. Under weakening overlap, it becomes a requirement.
\end{remark}

\begin{lemma}[Sufficient Condition for Variance Stability]\label{lem:var_stability_suff}
Suppose $\E[D^4] < \infty$ and let $\widehat{\sigma}_V^2 := n^{-1}\sum_{i=1}^n (D_i - \mhat^{(-k(i))}(X_i))^2$ denote the cross-fitted residual second moment. If
\[
\|\mhat^{(-k)} - m_0\|_{L^2} = o_P(\sigma_{V,n}) \quad \text{uniformly over } k,
\]
then $\widehat{\sigma}_V^2/\sigma_{V,n}^2 \pto 1$, hence Assumption~\ref{ass:var_stability} holds.

\end{lemma} 

\begin{proof}
Write $\Vhat_i = V_i + \Delta_i^m$. Then
\[
n^{-1}\sum_i \Vhat_i^2 = n^{-1}\sum_i V_i^2 + 2n^{-1}\sum_i V_i \Delta_i^m + n^{-1}\sum_i (\Delta_i^m)^2.
\]
By Cauchy--Schwarz, $|n^{-1}\sum_i V_i \Delta_i^m| \leq \|V\|_n \|\Delta^m\|_n = O_P(\sigma_{V,n}) \cdot o_P(\sigma_{V,n}) = o_P(\sigma_{V,n}^2)$. Also $n^{-1}\sum_i (\Delta_i^m)^2 = \|\Delta^m\|_n^2 = o_P(\sigma_{V,n}^2)$. Finally, $n^{-1}\sum_i V_i^2 \pto \sigma_{V,n}^2$ under finite fourth moments. \qed
\end{proof}

\begin{remark}[Interpretation of Sufficient Condition]\label{rem:var_stability_suff}
Lemma~\ref{lem:var_stability_suff} shows that, under weakening overlap, variance stability is ensured when the first-stage error is small relative to the residual scale $\sigma_{V,n}$. This prevents $\widehat{\sigma}_V^2$ from collapsing due to first-stage error.
\end{remark}

\begin{lemma}[Foldwise Empirical-to-Population Norm]\label{lem:emp_pop}
Let $\{I_1, \ldots, I_K\}$ be the cross-fitting folds (Definition~\ref{def:crossfit}). For each fold $k$, let $\Delta^{(-k)}: \mathcal{X} \to \R$ be any (possibly random) function measurable with respect to the training sigma-field generated by $\{W_j : j \notin I_k\}$. Define the fold empirical norm
\[
\|\Delta^{(-k)}\|_{n,k}^2 := \frac{1}{|I_k|}\sum_{i \in I_k} \Delta^{(-k)}(X_i)^2.
\]
Then, conditional on the training sample for fold $k$,
\[
\E\!\left[\|\Delta^{(-k)}\|_{n,k}^2 \mid \{W_j : j \notin I_k\}\right] = \|\Delta^{(-k)}\|_{L^2(P_X)}^2.
\]
Consequently, by Markov's inequality, $\|\Delta^{(-k)}\|_{n,k} = O_P(\|\Delta^{(-k)}\|_{L^2})$ uniformly over $k$.

If $\Delta_i := \Delta^{(-k)}(X_i)$ for $i \in I_k$, then the full-sample empirical norm satisfies
\[
\|\Delta\|_n^2 = \frac{1}{n}\sum_{k=1}^K |I_k|\, \|\Delta^{(-k)}\|_{n,k}^2,
\]
so in particular $\|\Delta\|_n = O_P\!\left(\max_{1 \leq k \leq K} \|\Delta^{(-k)}\|_{L^2}\right)$.
\end{lemma}

\begin{theorem}[Stochastic-order bound]\label{thm:fs_bound}
Under Assumptions~\ref{ass:causal_plr}, \ref{ass:moments}, \ref{ass:moment_bounds},
\ref{ass:rates}, \ref{ass:var_stability}, and \ref{ass:bounded_D}:
\begin{equation}\label{eq:fs_bound}
\thetahat - \theta_0
= O_P\!\left(\frac{\sqrt{\kappa_n}}{\sqrt{n}} + \kappa_n \cdot \mathrm{Rem}_n\right).
\end{equation}
where the remainder term is:
\begin{equation}\label{eq:remainder}
\mathrm{Rem}_n := r_n^m r_n^\ell + (r_n^m)^2 + \frac{r_n^m + r_n^\ell}{\sqrt{n}}.
\end{equation}
The oracle term $O_P(\sqrt{\kappa_n}/\sqrt{n})$ follows from $O_P(1/(\sigma_{V,n}\sqrt{n}))$ and Assumption~\ref{ass:bounded_D}.
\end{theorem}

\begin{proof}
From Theorem~\ref{thm:exact_decomp}: $\thetahat - \theta_0 = \widehat{\kappa}(S_n' + B_n')$.

Oracle term: The oracle term $S_n = n^{-1}\sum_i V_i\varepsilon_i$ has conditional mean zero and
\[
\Var(S_n) = \frac{1}{n}\E[V^2\varepsilon^2].
\]
By Assumption~\ref{ass:moment_bounds}(ii) and iterated expectations:
\[
\E[V^2\varepsilon^2] = \E[V^2 \E[\varepsilon^2 \mid D, X]] \leq C \cdot \E[V^2] = C \sigma_{V,n}^2.
\]
Thus $S_n = O_P(\sigma_{V,n}/\sqrt{n})$. By Assumption~\ref{ass:var_stability}, $\sigma_{V,n}^2/\widehat{\sigma}_V^2 = O_P(1)$, so:
\[
\widehat{\kappa} S_n' = \frac{S_n}{\widehat{\sigma}_V^2} = O_P\!\left(\frac{\sigma_{V,n}}{\sqrt{n}} \cdot \frac{1}{\sigma_{V,n}^2}\right) = O_P\!\left(\frac{1}{\sigma_{V,n}\sqrt{n}}\right).
\]

Bias term: By Lemma~\ref{lem:bias_decomp}, $B_n = \sum_{j=1}^5 B_n^{(j)}$.

Terms $B_n^{(1)}$--$B_n^{(3)}$: Under cross-fitting, these have conditional mean zero. For $B_n^{(1)} = n^{-1}\sum_i V_i \Delta_i^\ell$, by Assumption~\ref{ass:moment_bounds}(i) and iterated expectations:
\[
\E[V^2 (\Delta^\ell)^2] = \E[(\Delta^\ell)^2 \E[V^2 \mid X]] \leq C \|\Delta^\ell\|_{L^2}^2 = O_P((r_n^\ell)^2).
\]
Thus $\Var(B_n^{(1)}) = O_P((r_n^\ell)^2/n)$ and $B_n^{(1)} = O_P(r_n^\ell/\sqrt{n})$. Similarly for $B_n^{(2)}, B_n^{(3)}$.

Term $B_n^{(4)}$: By sample Cauchy--Schwarz:
\[
|B_n^{(4)}| = \left|\frac{1}{n}\sum_i \Delta_i^m \Delta_i^\ell\right| \leq \|\Delta^m\|_n \|\Delta^\ell\|_n.
\]
By Lemma~\ref{lem:emp_pop} applied foldwise to $\Delta^{m,(-k)}(x) := m_0(x) - \mhat^{(-k)}(x)$ and $\Delta^{\ell,(-k)}(x) := \ell_0(x) - \ellhat^{(-k)}(x)$, we have $\|\Delta^m\|_n = O_P(r_n^m)$ and $\|\Delta^\ell\|_n = O_P(r_n^\ell)$. Hence $|B_n^{(4)}| = O_P(r_n^m r_n^\ell)$.

Term $B_n^{(5)}$: Similarly, $|B_n^{(5)}| = |\theta_0| \|\Delta^m\|_n^2 = O_P((r_n^m)^2)$.

Combining: $B_n = O_P(r_n^m r_n^\ell + (r_n^m)^2 + (r_n^m + r_n^\ell)/\sqrt{n}) = O_P(\mathrm{Rem}_n)$.

Therefore: $\widehat{\kappa} B_n' = O_P(\kappa_n \cdot \mathrm{Rem}_n)$. \qedhere
\end{proof}

\begin{remark}[Complete Remainder]\label{rem:complete_rem}
The remainder~\eqref{eq:remainder} includes $(r_n^m)^2$ from $B_n^{(5)}$, which can dominate $r_n^m r_n^\ell$ when $r_n^m \gg r_n^\ell$. The cross-fitting terms contribute $(r_n^m + r_n^\ell)/\sqrt{n}$, negligible when $r_n^m, r_n^\ell = o(1)$.
\end{remark}

\begin{corollary}[Critical Rate]\label{cor:critical}
Under Assumption~\ref{ass:overlap} (strong overlap) and Assumption~\ref{ass:bounded_D}, a sufficient condition for valid $\sqrt{n}$-inference beyond oracle noise is
$\mathrm{Rem}_n = o(n^{-1/2})$, equivalently $\kappa_n \cdot \mathrm{Rem}_n = o(n^{-1/2})$.
\end{corollary}

\subsection{Efficiency Bound Connection}
\label{subsec:efficiency}

\begin{assumption}[Conditional Homoskedasticity]\label{ass:homosked}
$\E[\varepsilon^2 \mid D, X] = \sigma_\varepsilon^2$ almost surely.
\end{assumption}

\begin{remark}[Why Condition on $(D,X)$]\label{rem:homosked}
Assumption~\ref{ass:homosked} (conditional homoskedasticity) is stronger than the second-moment condition
$\E[\varepsilon^2\mid X]<\infty$. It is imposed here only to obtain a clean link between conditioning and the
efficiency bound, since it delivers the identity $\E[V^2\varepsilon^2]=\sigma_\varepsilon^2\,\sigma_V^2$.
\end{remark}

\begin{theorem}[Efficiency and Condition Number]\label{thm:efficiency}
Under Assumption~\ref{ass:homosked}, the semiparametric efficiency bound is:
\begin{equation}
V_{\mathrm{eff}} = \frac{\sigma_\varepsilon^2}{\sigma_V^2} = \frac{\sigma_\varepsilon^2 \kappa}{\sigma_D^2}.
\end{equation}
Without Assumption~\ref{ass:homosked}, the bound is $V_{\mathrm{eff}} = \E[V^2\varepsilon^2]/\sigma_V^4$. This efficiency result connects to the classical semiparametric variance bound literature \citep{newey1990semipar}.
\end{theorem}

\begin{proof}
The efficiency bound for moment $\E[\psi] = 0$ is $V_{\mathrm{eff}} = \E[\psi^2]/(\partial_\theta \E[\psi])^2$.

For $\psi = V\varepsilon$: $\E[\psi^2] = \E[V^2\varepsilon^2]$ and $\partial_\theta \E[\psi] = -\sigma_V^2$.

Under Assumption~\ref{ass:homosked}:
\begin{align*}
\E[V^2\varepsilon^2] &= \E[\E[V^2\varepsilon^2 \mid D, X]] \\
&= \E[V^2 \E[\varepsilon^2 \mid D, X]] \quad \text{($V^2$ is $(D,X)$-measurable)}\\
&= \E[V^2 \sigma_\varepsilon^2] = \sigma_\varepsilon^2 \sigma_V^2.
\end{align*}
Therefore: $V_{\mathrm{eff}} = \sigma_\varepsilon^2 \sigma_V^2 / \sigma_V^4 = \sigma_\varepsilon^2/\sigma_V^2 = \sigma_\varepsilon^2 \kappa/\sigma_D^2$. \qedhere
\end{proof}

\section{Conditioning Regimes and Rate Implications}
\label{sec:regimes}

The regimes below classify how overlap weakens along the triangular array. Plugging regime-specific $\kappa_n$ growth into Theorem~\ref{thm:fs_bound} yields the corresponding rate implications. Proofs for the regime-specific rate statements are provided in Appendix~A.3. From this point, we analyze sequences of DGPs under Assumption~\ref{ass:triangular}, replacing Assumption~\ref{ass:overlap} with $\Var(D_n \mid X_n = x) \geq \underline{\sigma}_n^2$ where $\underline{\sigma}_n^2 \downarrow 0$.

\begin{definition}[Conditioning Regimes]\label{def:regimes}
\leavevmode\par\vspace{0\baselineskip}
\begin{enumerate}[label=\textup{(\roman*)}, leftmargin=*]
\item \textbf{Well-conditioned:} $\kappa_n = O(1)$. Standard~$\sqrt{n}$\nobreakdash-asymptotics apply.
\item \textbf{Moderately ill-conditioned:} $\kappa_n = O(n^\gamma)$, $0<\gamma<1/2$. Slower convergence rates.
\item \textbf{Severely ill-conditioned:} $\kappa_n \asymp \sqrt{n}$. Standard~$\sqrt{n}$\nobreakdash-asymptotics fail.
\end{enumerate}
\end{definition}

\begin{theorem}[Rates by Regime]\label{thm:rates}
Suppose $\mathrm{Rem}_n = O(n^{-\alpha})$ and $\sigma_{D,n}^2 \asymp 1$:
\begin{enumerate}[label=(\roman*)]
\item Well-conditioned: $\thetahat - \theta_0 = O_P(n^{-1/2})$.
\item Moderately ill-conditioned: $\thetahat - \theta_0 = O_P(n^{\gamma/2 - 1/2} + n^{\gamma-\alpha})$.
\item Severely ill-conditioned: $\thetahat - \theta_0 = O_P(n^{-1/4} + n^{1/2-\alpha})$.
\end{enumerate}
\end{theorem}

We interpret $\kappaOof$ through qualitative conditioning regimes that guide reporting practices rather than impose strict cutoffs. In well-conditioned scenarios, the orthogonal score exhibits sufficient curvature to ensure that the estimating equation is well-posed. In this case, residual nuisance error is not substantially amplified. As conditioning deteriorates, the same nuisance error becomes increasingly magnified. Consequently, the sensitivity of cross-learner and nuisance specifications serves as direct, observable indicator of estimator fragility and should be reported. In severely ill-conditioned contexts, the sampling term may remain small while an amplified systematic component dominates the estimator. Under such circumstances, conventional confidence intervals tend to under-cover.

A recommended workflow involves computing $\kappaOof$ for each learner and reporting these values alongside a multi-learner sensitivity summary, as demonstrated in the empirical illustration. Notably, $\kappaOof$ functions as a diagnostic for the conditioning of the orthogonalized score, rather than as evidence for unconfoundedness or causal validity. When $\kappaOof$ is elevated, estimates should be regarded as fragile. Selection of a single preferred learner should be avoided. Overlap adjustments represent natural complementary responses in applied research, although a formal analysis of these remedies is beyond the scope of this paper.

\section{Monte Carlo Evidence}
\label{sec:simulations}

We validate the theoretical predictions via a corrupted oracle design that isolates the amplification mechanism. Predictions P1--P3 follow from Theorems~\ref{thm:exact_decomp}--\ref{thm:fs_bound}:
\begin{description}[leftmargin=1em,labelindent=0em]
\item[\textbf{P1}] For fixed nuisance-error $\delta$, $|\text{Bias}|$ increases proportionally with $\kappa$.
\item[\textbf{P2}] $|\text{Bias}|/\text{SE}$ is approximately a function of the single index $\kappa \times \delta$.
\item[\textbf{P3}] Coverage fails when $|\text{Bias}|/\text{SE}$ increases 
\end{description}

We generate $n = 2{,}000$ observations from the PLR model with $\theta_0 = 1$, $X \in \mathbb{R}^{10}$ drawn from $N(0, \Sigma)$ with AR(1) correlation $\Sigma_{jk} = 0.5^{|j-k|}$ and $U, \varepsilon \stackrel{\text{i.i.d.}}{\sim} N(0, 1)$. The treatment equation is $D = m_0(X) + \sigma_U U$ where $m_0(X) = \beta^\top X$ with $\beta_j = 0.7^j$ and $\sigma_U$ calibrated to target $R^2(D|X) \in {0.50, 0.75, 0.90, 0.95, 0.97, 0.99}$, yielding $\kappa \in {2, 4, 10, 20, 33, 100}$. The corrupted oracle injects multiplicative bias: $\widehat{m}(X) = m_0(X)(1 + \delta)$ and $\widehat{\ell}(X) = \ell_0(X)(1 + \delta)$ for $\delta \in {0, 0.02, 0.05, 0.10, 0.20}$.

\begin{figure}[htbp]
\centering
\maybeincludegraphics[width=\textwidth]{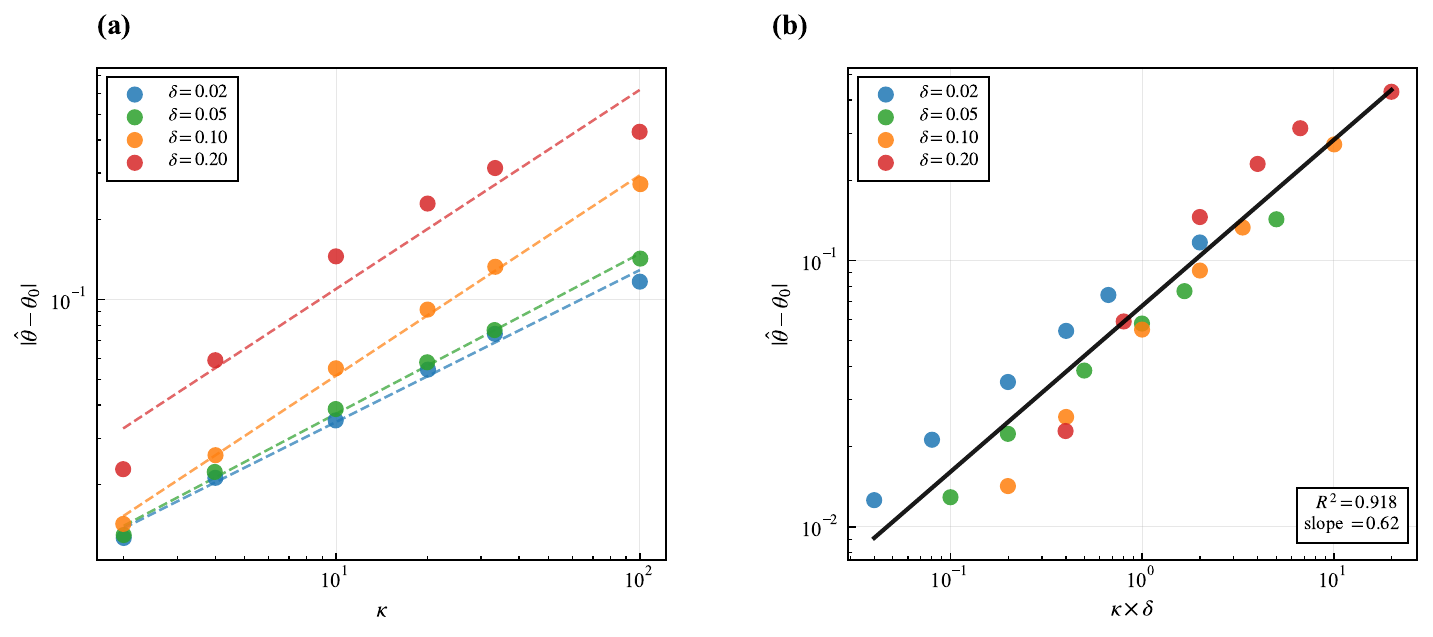}
\caption{Bias amplification mechanism ($n = 2{,}000$, $B = 500$). (a) Mean absolute bias vs.\ structural $\kappa$ for each $\delta \in \{0.02, 0.05, 0.10, 0.20\}$. (b) All points collapse onto a single trend when plotted against $\kappa \times \delta$.}
\label{fig:bias}
\end{figure}

Figure~\ref{fig:bias} visualizes the bias-amplification mechanism. For fixed nuisance error magnitude, increasing conditioning (higher $\kappa$) magnifies the resulting estimation error. Panel (a) confirms Prediction P1. For each fixed $\delta$, bias increases monotonically with $\kappa$, and the approximately parallel lines across $\delta$ levels illustrate the multiplicative structure. Each unit increase in $\log \kappa$ shifts $\log|\text{Bias}|$ by roughly the same amount regardless of $\delta$. Panel (b) confirms Prediction P2. When bias is plotted against the single index $\kappa\times\delta$, the points align along a single monotone trend, and the pooled log--log regression
\begin{equation}\label{eq:loglog}
\log|\thetahat-\theta_0|=-2.69+0.62\cdot\log(\kappa\times\delta),\quad R^2=0.92,
\end{equation}
captures this organization. The estimated slope below one is consistent with the finite-sample theory. The remainder term in Theorem~\ref{thm:fs_bound} contains both linear and quadratic components (e.g., $r_n^m r_n^\ell+(r_n^m)^2$), and the presence of a nonzero noise floor at $\delta=0$ mechanically flattens the log--log relationship for small values of $\kappa\times\delta$. Table~\ref{tab:exponent} further reports separate elasticities, reinforcing the finite-sample amplification mechanism.

\begin{figure}[htbp]
\centering
\maybeincludegraphics[width=\textwidth]{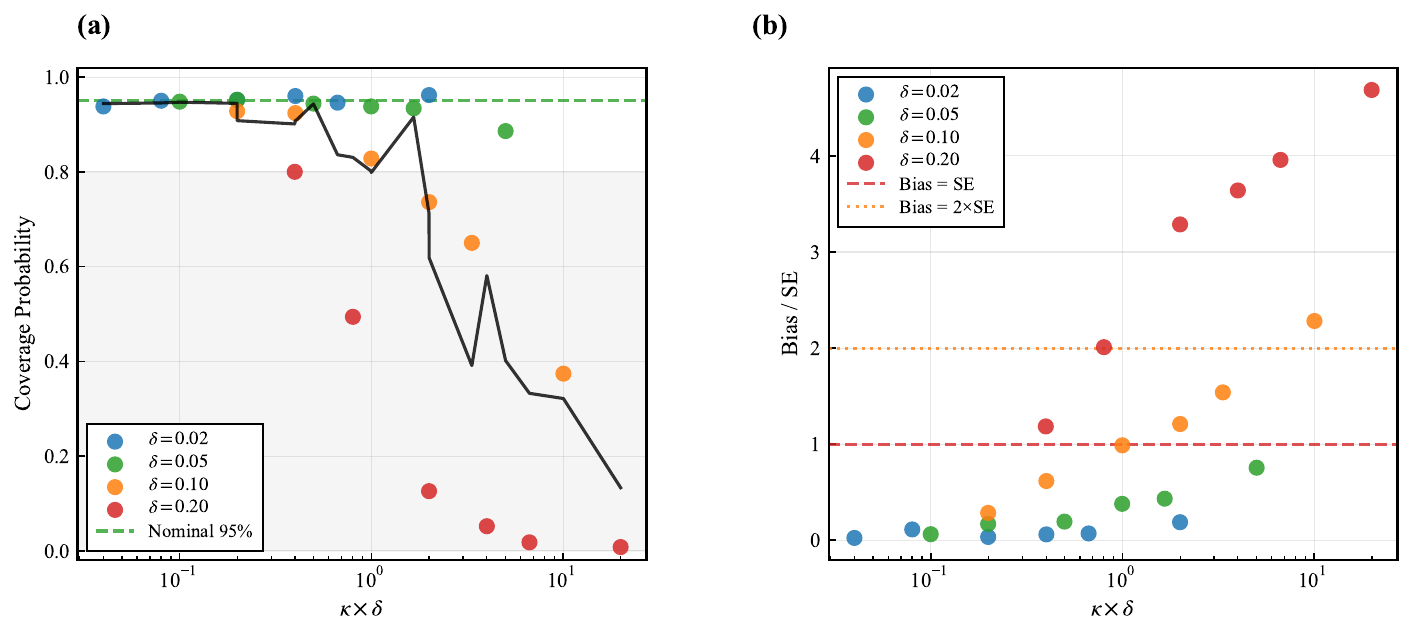}
\caption{Coverage mechanism ($n = 2{,}000$, $B = 500$). (a) Coverage vs.\ $\kappa \times \delta$: confirms single-index organization. (b) Coverage vs.\ $|\text{Bias}|/\text{SE}$: the horizontal dashed line at 1.0 corresponds to Bias = SE and marks the transition---normal-approximation CIs fail when bias dominates noise.}
\label{fig:coverage}
\end{figure}

Figure~\ref{fig:coverage} shows that coverage remains near nominal in well-conditioned regimes, but collapses sharply as $\kappa$ increases for fixed $\delta$, consistent with P3. At $\delta = 0$, coverage ranges from 0.94 to 0.97 across $\kappa$ (see Table~\ref{tab:coverage} for the full grid). Table~\ref{tab:kappa} reports the structural $\kappa$ values associated with each $R^2$ regime and confirms $\kappa$ is invariant to injected bias. Coverage decreases primarily because $|\text{Bias}|/\text{SE}$ increases with $\kappa \times \delta$, not because SE inflates. This is a silent failure where CIs are narrow but miss $\theta_0$.

Panel (b) reveals why coverage fails: the ratio $|\text{Bias}|/\text{SE}$ crosses 1.0 precisely when coverage drops below 80\%. This confirms that undercoverage is due to bias amplification, not variance inflation. Sample-size sensitivity is summarized in Table~\ref{tab:samplesize}.

The Monte Carlo evidence confirms all three predictions. A sign-structure experiment in Table~\ref{tab:opposite} validates the cross-term structure of Lemma~\ref{lem:bias_decomp}. Opposite-sign nuisance errors produce $3.7\times$ larger bias and near-zero coverage at high $\kappa$, confirming that asymmetric regularization is more damaging than symmetric bias. Robustness to nonlinear DGPs appears in Table~\ref{tab:nonlinear}.

When $\kappa$ is large, small systematic nuisance errors can dominate sampling noise. In this case, bias is amplified beyond what standard errors capture. The single-index $\kappa \times \delta$ organizes when coverage collapses, making $\kappa$ a fragility predictor. Instability across learner implementations is itself informative about estimation sensitivity. For practice, the main implication is reporting-oriented rather than threshold-based. Large estimated $\kappa$ should be interpreted as a fragility flag that motivates transparent sensitivity analysis.

\section{Empirical Application}
\label{sec:application}

We demonstrate our proposed reporting workflow in the LaLonde \citep{lalonde1986} benchmark, using the NSW experimental
sample and the NSW--PSID observational comparison design studied by \citet{dehejiawahba1999}. The purpose of this section
is diagnostic and interpretation rather than causal adjudication. In particular, the observational benchmark cannot disentangle implementation fragility induced by weak residualized treatment variation (a conditioning problem)
from genuine violations of causal assumptions (an identification problem). Our goal is instead to show how the out-of-fold
conditioning diagnostic $\hat\kappa_{\mathrm{oof}}$ behaves in a standard applied PLR--DML pipeline. Throughout we use $K=5$ cross-fitting and run PLR--DML across six learners (OLS, Lasso, Ridge, RF, GBM, MLP).\footnote{
For simplicity, we use the same learner class for both $\widehat m(X)$ and $\widehat \ell(X)$. Mixed learner choices are
straightforward but not pursued here.}

\begin{table}[!tb]
\small
\centering
\caption{LaLonde baseline PLR--DML estimates by sample and learner.}
\label{tab:lalonde_baseline_main}
\setlength{\tabcolsep}{4.5pt} % tighter columns
\renewcommand{\arraystretch}{1.05} % slightly tighter rows
\begin{tabular}{l l r r r r r r r}
\toprule
Sample & Learner & Estimate & SE & CI\_Lower & CI\_Upper & $\hat R^2_{\mathrm{oof}}(D\mid X)$ & $\hat\kappa_{\mathrm{oof}}$ & $N$ \\
\midrule
Experimental & OLS   & 1{,}598 & 668 &   290 & 2{,}907 & 0.000 & 1.00 &   445 \\
Experimental & Lasso & 1{,}700 & 673 &   381 & 3{,}020 & 0.000 & 1.00 &   445 \\
Experimental & Ridge & 1{,}615 & 666 &   310 & 2{,}921 & 0.000 & 1.00 &   445 \\
Experimental & RF    & 1{,}562 & 651 &   287 & 2{,}838 & 0.000 & 1.00 &   445 \\
Experimental & GBM   & 1{,}810 & 662 &   511 & 3{,}108 & 0.000 & 1.00 &   445 \\
Experimental & MLP   & 1{,}019 & 847 &  $-642$ & 2{,}680 & 0.000 & 1.00 &   445 \\
\midrule
Observational & OLS   &   700 & 785 &  $-839$  & 2{,}239 & 0.291 & 1.41 & 2{,}675 \\
Observational & Lasso &   190 & 634 & $-1{,}053$ & 1{,}432 & 0.010 & 1.01 & 2{,}675 \\
Observational & Ridge &   699 & 784 &  $-837$  & 2{,}236 & 0.291 & 1.41 & 2{,}675 \\
Observational & RF    &  $-803$ & 969 & $-2{,}702$ & 1{,}096 & 0.611 & 2.57 & 2{,}675 \\
Observational & GBM   & $-1{,}158$ & 957 & $-3{,}033$ &   717 & 0.618 & 2.62 & 2{,}675 \\
Observational & MLP   &  $-776$ & 939 & $-2{,}616$ & 1{,}064 & 0.605 & 2.53 & 2{,}675 \\
\bottomrule
\end{tabular}

\medskip
{\footnotesize
\begin{minipage}{0.97\linewidth}
\emph{Notes:} $K=5$ cross-fitting. $\hat R^2_{\mathrm{oof}}(D\mid X)$ is the out-of-fold first-stage fit for predicting $D$ from $X$
using the learner indicated. The diagnostic is $\hat\kappa_{\mathrm{oof}} = 1/(1-\hat R^2_+)$ with
$\hat R^2_+=\max\{0,\,\hat R^2_{\mathrm{oof}}(D\mid X)\}$.
\end{minipage}
}
\end{table}

Table~\ref{tab:lalonde_baseline_main} reports $\hat\theta$, standard errors, confidence intervals, the out-of-fold
first-stage fit $\hat R^2_{\mathrm{oof}}(D\mid X)$ and the implied conditioning diagnostic
$\hat\kappa_{\mathrm{oof}} = 1/(1-\hat R^2_+)$ with $\hat R^2_+=\max\{0,\hat R^2_{\mathrm{oof}}(D\mid X)\}$. In the experimental NSW sample,
$\hat R^2_{\mathrm{oof}}(D\mid X)$ is near zero, so $\hat\kappa_{\mathrm{oof}}\approx 1$ for all learners and estimates are comparatively
stable across nuisance specifications. In the observational NSW--PSID sample, more flexible learners achieve higher out-of-fold first-stage fit, implying larger $\hat\kappa_{\mathrm{oof}}$.
These same specifications exhibit some larger cross-learner dispersion, including sign reversals, despite confidence
interval widths that do not mechanically expand one-for-one with dispersion. This is the qualitative empirical pattern
predicted by our exact PLR identity and bound.\footnote{In many observational settings $\hat R^2_{\mathrm{oof}}(D\mid X)$ can be materially higher than in Table~\ref{tab:lalonde_baseline_main}, so $\hat\kappa_{\mathrm{oof}}$ may be much larger and the amplification mechanism correspondingly more pronounced. We use the LaLonde benchmark because it admits comparison to an experimental benchmark estimate. As noted above, the observational comparison does not disentangle conditioning-driven fragility from identification failures.} As residualized treatment variation weakens, amplification can make otherwise
second-order implementation differences (learner class, tuning, splitting, and regularization) economically consequential
for $\hat\theta$.

\begin{figure}[!tb]
\centering
\maybeincludegraphics[width=0.95\textwidth]{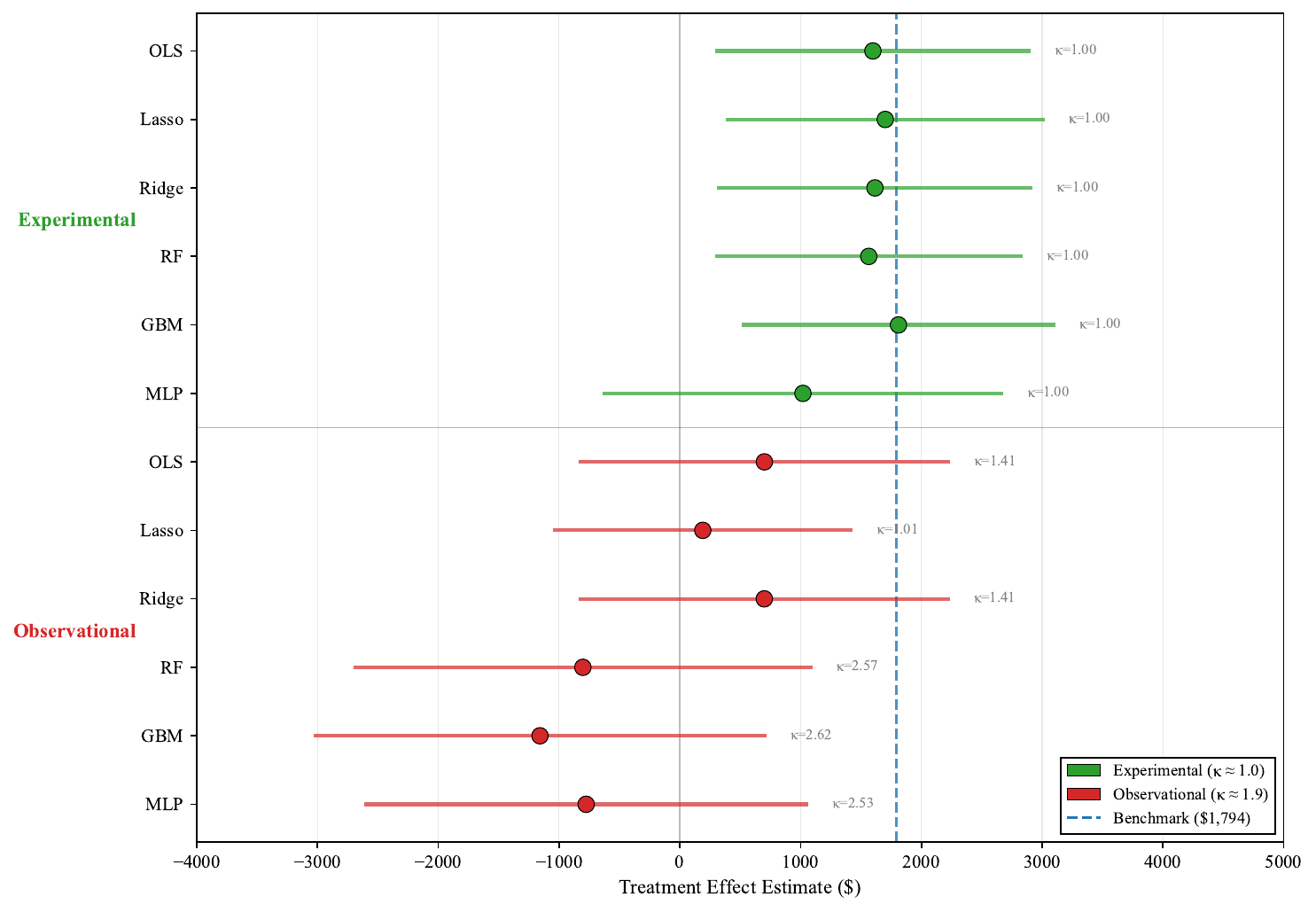}
\caption{Forest plot of PLR--DML estimates by learner and sample, reported alongside $\hat\kappa_{\mathrm{oof}}$ as a fragility diagnostic.
Experimental: tight clustering with $\hat\kappa_{\mathrm{oof}}\approx 1$. Observational: larger dispersion and occasional sign reversals
as $\hat\kappa_{\mathrm{oof}}$ rises in more flexible first-stage specifications. The benchmark line corresponds to the experimental benchmark estimate reported in LaLonde (1986).}
\label{fig:lalonde_forest}
\end{figure}

Accordingly, we interpret $\hat\kappa_{\mathrm{oof}}$ narrowly as a learner-specific proxy for orthogonal-score conditioning. Our recommendation is procedural and reporting-oriented:
(a) pre-specify a menu of reasonable nuisance learners and tuning rules (including cross-fitting and tuning conventions);
(b) compute $\hat\kappa_{\mathrm{oof}}$ learner-by-learner from the same out-of-fold residuals that enter the cross-fitted score; and
(c) report $\hat\kappa_{\mathrm{oof}}$ together with a compact multi-learner sensitivity summary, such as a forest plot (see Figure~\ref{fig:lalonde_forest}). When $\hat\kappa_{\mathrm{oof}}$ is elevated and dispersion is large, nominal standard
errors can understate uncertainty because amplification may dominate sampling noise.

\section{Conclusion}
\label{sec:conclusion}

We show that, in PLR-DML, conditioning of the orthogonal score is a first-order amplification channel for finite-sample error. The exact identity $\hat\theta-\theta_0=\widehat{\kappa}(S_n' + B_n')$ makes the multiplicative role of the condition number explicit, and the stochastic bound $\hat\theta-\theta_0 = O_P(\sqrt{\kappa_n/n}+\kappa_n\mathrm{Rem}_n)$ yields an operational sufficiency condition: $\kappa_n\mathrm{Rem}_n=o(n^{-1/2})$. For practice, we recommend reporting $\hat\kappa_{\mathrm{oof}}:=1/\{1-\max(0,\hat R^2_{\mathrm{oof}}(D\mid X))\}$ alongside estimates. A large $\hat\kappa_{\mathrm{oof}}$ indicates fragility. In this case, narrow CIs should be interpreted cautiously and multi-learner sensitivity summaries are recommended. 

Our results are exact for PLR. In more general orthogonal-score problems, the diagnostic framework extends by linearizing the score equation and tracking both (i) the score Jacobian (the stability object) and (ii) the associated linearization remainder. Appendix~A.1 provides the generic orthogonal-score identity that anchors this template. Developing weak-conditioning-robust inference procedures, and extending the analysis to heterogeneous treatment effects
and other orthogonal-score settings, are promising directions.

\section*{Disclosure Statement}
The author has no conflicts of interest to declare.

\section*{Data Availability Statement}
Replication code and data are available at \url{https://github.com/gsaco/dml-diagnostic}.

\section*{Acknowledgments}
I used Grammarly for language editing. All remaining errors are my own.

\newpage

\bibliographystyle{apalike}

%%%%%%%%%%%%%%%%%%%%%%%%%%%%%%%%%%%%%%%%%%%%%%%%%%%%%%%%%%%%%%%%%%%%%%%%%%%%%%%
% APPENDIX
%%%%%%%%%%%%%%%%%%%%%%%%%%%%%%%%%%%%%%%%%%%%%%%%%%%%%%%%%%%%%%%%%%%%%%%%%%%%%%%

\newpage
\appendix

% Reset and renumber tables/figures for appendix
\renewcommand{\thetable}{A.\arabic{table}}
\setcounter{table}{0}
\renewcommand{\thefigure}{A.\arabic{figure}}
\setcounter{figure}{0}

\section{Mathematical Appendix}
\label{app:proofs}

\noindent All notation follows Section 3 of the main paper. We restate only what is needed.

\subsection{A Generic Orthogonal-Score Identity}
\label{app:generic_identity}

This paper uses a special feature of PLR: the orthogonal score is affine in $\theta$, so the
DML estimating equation can be solved exactly and yields an exact decomposition without any
Taylor/von Mises remainder.
For completeness, we record the corresponding generic identity for cross-fitted score estimators,
which highlights the same ``Jacobian amplification'' channel in models where the score need not be
affine in $\theta$.

\textbf{Setup.} Let $\psi(W;\theta,\eta)\in\mathbb{R}^{d_\theta}$ be a (possibly nonlinear) score, where $\theta\in\Theta\subset\mathbb{R}^{d_\theta}$
and $\eta$ is a nuisance element. Using a $K$-fold cross-fitting structure, let $\hat\eta^{(-k)}$ denote the nuisance estimator trained on $\{W_j: j\notin I_k\}$.
Define the cross-fitted empirical moment map
\begin{equation}
\label{eq:generic_Psi_n}
\Psi_n(\theta,\hat\eta)
:=
\frac{1}{n}\sum_{k=1}^K \sum_{i\in I_k}
\psi\!\left(W_i;\theta,\hat\eta^{(-k)}\right).
\end{equation}
Let $\hat\theta$ be any solution to $\Psi_n(\hat\theta,\hat\eta)=0$.

\textbf{Mean-value (Jacobian) identity.}
Assume that $\theta\mapsto\Psi_n(\theta,\hat\eta)$ is continuously differentiable on the line segment
$\{\theta_0+t(\hat\theta-\theta_0): t\in[0,1]\}$ and define the intermediate Jacobian
\begin{equation}
\label{eq:J_e_def}
\widehat J_n^{e}
:=
\int_{0}^{1}
\partial_{\theta}\Psi_n\!\left(\theta_0+t(\hat\theta-\theta_0),\hat\eta\right)\,dt.
\end{equation}
Then the vector mean-value theorem implies
\begin{equation}
\label{eq:generic_mvt_identity}
0=\Psi_n(\hat\theta,\hat\eta)=\Psi_n(\theta_0,\hat\eta)+\widehat J_n^{e}(\hat\theta-\theta_0).
\end{equation}
If $\widehat J_n^{e}$ is nonsingular, rearranging yields the identity
\begin{equation}
\label{eq:generic_theta_identity}
\hat\theta-\theta_0
=
-(\widehat J_n^{e})^{-1}\,\Psi_n(\theta_0,\hat\eta).
\end{equation}

\textbf{Oracle--bias decomposition.}
Add and subtract the oracle score:
\begin{equation}
\label{eq:generic_S_B}
\Psi_n(\theta_0,\hat\eta)
=
\underbrace{\Psi_n(\theta_0,\eta_0)}_{=:S_n^{\mathrm{gen}}}
+
\underbrace{\big\{\Psi_n(\theta_0,\hat\eta)-\Psi_n(\theta_0,\eta_0)\big\}}_{=:B_n^{\mathrm{gen}}}.
\end{equation}
Combining yields the decomposition
\begin{equation}
\label{eq:generic_decomp}
\hat\theta-\theta_0
=
-(\widehat J_n^{e})^{-1}\Big(S_n^{\mathrm{gen}}+B_n^{\mathrm{gen}}\Big).
\end{equation}

In PLR, the orthogonal score is affine in $\theta$, so
$\partial_{\theta}\Psi_n(\theta,\hat\eta)$ does not depend on $\theta$, and the intermediate Jacobian
$\widehat J_n^{e}$ collapses to the empirical Jacobian used in the main text.
Consequently the generic decomposition reduces to the exact identity in the main paper,
with no linearization remainder.

For nonlinear scores, the decomposition remains an exact algebraic identity conditional on
differentiability (it is simply the mean-value theorem applied to the sample estimating equation),
but controlling $B_n^{\mathrm{gen}}$ and replacing $(\widehat J_n^{e})^{-1}$ by a convenient plug-in
typically requires additional smoothness and orthogonality arguments, and yields explicit remainder terms.

\subsection{Proof of Neyman Orthogonality}

\begin{proof}
Let $\eta_r = \eta_0 + r(\eta - \eta_0) = (\ell_r, m_r)$. Define $\tilde{V}_r := D - m_r(X)$, $\tilde{U}_r := Y - \ell_r(X)$.

The score: $\psi(W; \theta_0, \eta_r) = \tilde{V}_r\{\tilde{U}_r - \theta_0 \tilde{V}_r\}$.

Taking the derivative at $r=0$:
\[
\left.\frac{\partial}{\partial r}\psi\right|_{r=0} = -(m-m_0)(X) \cdot \varepsilon + V \cdot \{-(\ell-\ell_0)(X) + \theta_0(m-m_0)(X)\}.
\]

Taking expectations, each term vanishes:
\begin{itemize}
\item By $\E[\varepsilon \mid X] = \E[\E[\varepsilon \mid D, X] \mid X] = 0$, hence $\E[(m-m_0)(X) \cdot \varepsilon] = \E[(m-m_0)(X) \cdot \E[\varepsilon \mid X]] = 0$.
\item $\E[V \cdot (\ell-\ell_0)(X)] = \E[\E[V \mid X] \cdot (\ell-\ell_0)(X)] = 0$.
\item $\E[V \cdot (m-m_0)(X)] = \E[\E[V \mid X] \cdot (m-m_0)(X)] = 0$, since $\E[V \mid X] = 0$. \qedhere
\end{itemize}
\end{proof}

\subsection{Proof of Rates by Regime}

\begin{proof}
Substitute regime-specific $\kappa_n$ into the stochastic-order bound. Assume $\sigma_{D,n}^2 \asymp 1$ so the oracle term is $O_P(\sqrt{\kappa_n}/\sqrt{n})$.

\textit{(i) Well-conditioned:} $\kappa_n = O(1)$ gives oracle $O_P(n^{-1/2})$ and bias $O_P(n^{-\alpha})$, so $O_P(n^{-1/2})$.

\textit{(ii) Moderately ill-conditioned:} $\kappa_n = O(n^\gamma)$ gives oracle $O_P(n^{\gamma/2}/\sqrt{n}) = O_P(n^{\gamma/2 - 1/2})$ and bias $O_P(n^{\gamma-\alpha})$.

\textit{(iii) Severely ill-conditioned:} $\kappa_n \asymp \sqrt{n}$ gives oracle $O_P(n^{1/4}/\sqrt{n}) = O_P(n^{-1/4})$ and bias $O_P(n^{1/2-\alpha})$. If $\alpha < 1/2$, bias diverges. \qedhere
\end{proof}

\newpage
\section{Simulation Appendix}
\label{app:simulations}

This section provides supporting tables for the Monte Carlo analysis in the main text.

\subsection{Coverage probability grid}

\begin{table}[htbp]
\centering
\caption{Coverage probability by nuisance bias $\delta$ and conditioning $R^2(D|X)$.}
\label{tab:coverage}
\begin{tabular}{l cccccc}
\toprule
& \multicolumn{6}{c}{$R^2(D|X)$ (structural $\kappa$)} \\
\cmidrule(lr){2-7}
$\delta$ & 0.50 (2) & 0.75 (4) & 0.90 (10) & 0.95 (20) & 0.97 (33) & 0.99 (100) \\
\midrule
0.00 & 0.94 & 0.95 & 0.94 & 0.95 & 0.97 & 0.94 \\
0.02 & 0.94 & 0.95 & 0.95 & 0.96 & 0.95 & 0.96 \\
0.05 & 0.95 & 0.95 & 0.94 & 0.94 & 0.93 & 0.89 \\
0.10 & 0.93 & 0.92 & 0.83 & \textbf{0.74} & \textbf{0.65} & \textbf{0.37} \\
0.20 & 0.80 & \textbf{0.49} & \textbf{0.13} & \textbf{0.05} & \textbf{0.02} & \textbf{0.01} \\
\bottomrule
\end{tabular}

\medskip
\begin{minipage}{\linewidth}
\footnotesize
\textbf{Notes:} Sample size $=2,000$, Monte Carlo replications $B=500$. Confidence intervals are based on the standard normal approximation using DML influence-function-based standard errors (HC-robust sandwich estimator). ``Coverage'' is the proportion of such 95\% CIs containing the true parameter $\theta_0=1$. Bias $\delta$ is injected multiplicatively: $\hat{\eta}(x) = \eta_0(x)(1+\delta)$.
\end{minipage}
\end{table}

\newpage
\subsection{Log--log exponent diagnostic for bias scaling}

To understand the log-log slope in the main text, we estimate separate exponents on $\kappa$ and $\delta$ via multivariate regression:
\begin{equation}\label{eq:exponent_reg}
\log|\text{Bias}| = a + \alpha \log \kappa + \beta \log \delta + \text{error}.
\end{equation}
We exclude $\delta=0$ rows since $\log(0)$ is undefined. Observations correspond to the aggregate mean absolute bias for each of the $30$ simulation grid cells (6 $R^2$ regimes $\times$ 5 $\delta$ levels), weighted equally.

\begin{table}[htbp]
\centering
\caption{Exponent diagnostic: $\log|\text{Bias}| \sim \alpha \log \kappa + \beta \log \delta$.}
\label{tab:exponent}
\begin{tabular}{l cc}
\toprule
Parameter & Estimate & Std.\ Error \\
\midrule
$\alpha$ (exponent on $\kappa$) & 0.67 & 0.06 \\
$\beta$ (exponent on $\delta$) & 0.51 & 0.05 \\
$R^2$ & \multicolumn{2}{c}{0.93} \\
\bottomrule
\end{tabular}

\medskip
\begin{minipage}{\linewidth}
\footnotesize
\textbf{Notes:} Multivariate OLS regression of $\log|\hat{\theta} - \theta_0|$ on $\log \kappa$ and $\log \delta$. The regression uses 30 observations, where each observation corresponds to the aggregate mean absolute bias for one simulation grid cell (6 $R^2$ regimes $\times$ 5 $\delta$ levels). Rows with $\delta=0$ are excluded as $\log(0)$ is undefined. Standard errors are classical OLS.
\end{minipage}
\end{table}

The stochastic-order bound implies that the bias component is of order $\kappa \mathrm{Rem}_n$, where
$\mathrm{Rem}_n$ collects products and squares of nuisance errors. We therefore report $(\hat\alpha,\hat\beta)$
as a descriptive summary of how Monte Carlo bias scales with $(\kappa,\delta)$ over the design grid,
not as a test of a sharp theoretical log--log slope in $\delta$.

\newpage
\subsection[Structural kappa stability]{Structural $\kappa$ Stability}

Table~\ref{tab:kappa} confirms that structural $\kappa$ (computed from true population residuals) is invariant to injected bias $\delta$.

\begin{table}[htbp]
\centering
\caption{Structural $\kappa$ by $R^2$ regime (rows) and bias level $\delta$ (columns).}
\label{tab:kappa}
\begin{tabular}{l ccccc}
\toprule
$R^2(D|X)$ & $\delta=0$ & $\delta=0.02$ & $\delta=0.05$ & $\delta=0.10$ & $\delta=0.20$ \\
\midrule
0.50 & 2.00 & 2.00 & 2.00 & 1.99 & 1.99 \\
0.75 & 3.98 & 4.01 & 3.99 & 4.01 & 4.00 \\
0.90 & 9.99 & 9.97 & 9.97 & 9.99 & 9.99 \\
0.95 & 20.03 & 20.06 & 19.94 & 20.00 & 20.00 \\
0.97 & 33.35 & 33.30 & 33.23 & 33.42 & 33.37 \\
0.99 & 100.08 & 99.90 & 100.24 & 100.34 & 99.74 \\
\bottomrule
\end{tabular}

\medskip
\begin{minipage}{\linewidth}
\footnotesize
\textbf{Notes:} Sample size $=2,000$. Values report the median structural $\kappa$ across $B=500$ replications.
\end{minipage}
\end{table}

This stability confirms that the corrupted oracle design correctly isolates the amplification mechanism: $\kappa$ is determined by the DGP alone, not by learner choice or injected bias.

\newpage
\subsection{Nonlinear DGP Robustness}

To ensure the mechanism is not an artifact of linear propensity, we replace the linear $m_0(X) = \beta^\top X$ with $m_0(X) = \tanh(\beta^\top X) \times c$, where $c$ is calibrated to match target $R^2$.

\begin{table}[htbp]
\centering
\caption{Linear vs.\ nonlinear DGP ($\delta = 0.1$, $B = 500$). The mechanism persists.}
\label{tab:nonlinear}
\begin{tabular}{l ccccc}
\toprule
DGP & $R^2$ & $\kappa$ & $|\text{Bias}|$ & Coverage & $|\text{Bias}|$/SE \\
\midrule
Linear & 0.75 & 4 & 0.026 & 0.92 & 0.92 \\
Nonlinear & 0.75 & 4 & 0.026 & 0.90 & 0.94 \\
Linear & 0.90 & 10 & 0.055 & 0.83 & 1.16 \\
Nonlinear & 0.90 & 10 & 0.049 & 0.86 & 1.03 \\
Linear & 0.95 & 20 & 0.092 & 0.74 & 1.38 \\
Nonlinear & 0.95 & 20 & 0.090 & 0.74 & 1.35 \\
\bottomrule
\end{tabular}

\medskip
\begin{minipage}{\linewidth}
\footnotesize
\textbf{Notes:} Sample size $=2,000$, $B=500$. Nonlinear DGP uses $m_0(X) = c \cdot \tanh(\beta^\top X)$.
\end{minipage}
\end{table}

Both DGPs exhibit bias amplification with similar magnitude. The qualitative conclusion is unchanged: the amplification mechanism is not an artifact of linearity.

\newpage
\subsection{Sample Size Sensitivity}

We examine whether larger $n$ resolves the problem by holding $R^2 = 0.90$ ($\kappa \approx 10$) and $\delta = 0.1$ fixed while varying $n \in \{500, 1000, 2000, 4000\}$. Bias decreases slowly with $n$, but SE decreases faster. The ratio $|\text{Bias}|/\text{SE}$ increases with $n$, causing coverage to worsen.

\begin{table}[htbp]
\centering
\caption{Sample size sensitivity ($R^2 = 0.90$, $\delta = 0.1$, $B = 500$).}
\label{tab:samplesize}
\begin{tabular}{l cccc}
\toprule
$n$ & $|\text{Bias}|$ & SE & $|\text{Bias}|$/SE & Coverage \\
\midrule
500 & 0.087 & 0.095 & 0.92 & 0.91 \\
1,000 & 0.064 & 0.067 & 0.96 & 0.90 \\
2,000 & 0.049 & 0.047 & 1.04 & 0.86 \\
4,000 & 0.048 & 0.033 & 1.45 & 0.73 \\
\bottomrule
\end{tabular}

\medskip
\begin{minipage}{\linewidth}
\footnotesize
\textbf{Notes:} $\delta=0.1$ fixed. Monte Carlo $B=500$. Shows persistence of bias amplification and coverage degradation as $n$ increases.
\end{minipage}
\end{table}

\newpage
\subsection{Sign-Structure Mechanism Check}
\label{app:sign}

Theory implies that cross-terms in the bias decomposition can cancel or reinforce depending on sign alignment of nuisance errors. Same-sign vs.\ opposite-sign is a designed falsification check: changing only sign structure should change amplification dramatically. ``Same-sign'' bias implies multiplicative bias $(1+\delta)$ for both nuisance functions. ``Opposite-sign'' bias implies scaling $m_0$ by $(1+\delta)$ and $\ell_0$ by $(1-\delta)$, removing the cancellation of product terms. Table~\ref{tab:opposite} confirms this prediction: opposite-sign nuisance errors maximize amplification and yield the most severe coverage failures.

\begin{table}[htbp]
\centering
\caption{Same-sign vs.\ opposite-sign bias.}
\label{tab:opposite}
\begin{tabular}{l cccccc}
\toprule
$R^2$ & $\kappa$ & Sign & $|\text{Bias}|$ & Coverage & Ratio vs.\ same \\
\midrule
0.75 & 4 & same & 0.027 & 0.90 & -- \\
0.75 & 4 & opposite & 0.072 & 0.32 & 2.7$\times$ \\
0.90 & 10 & same & 0.052 & 0.84 & -- \\
0.90 & 10 & opposite & 0.210 & 0.00 & 4.0$\times$ \\
0.95 & 20 & same & 0.091 & 0.75 & -- \\
0.95 & 20 & opposite & 0.408 & 0.00 & 4.5$\times$ \\
\midrule
\multicolumn{5}{l}{Average amplification ratio} & 3.7$\times$ \\
\bottomrule
\end{tabular}

\medskip
\begin{minipage}{\linewidth}
\footnotesize
\textbf{Notes:} Sample size $=2,000$, $B=500$. ``Same-sign'' injection: $\hat{m} = m_0(1+\delta)$ and $\hat{\ell} = \ell_0(1+\delta)$. ``Opposite-sign'' injection: $\hat{m} = m_0(1+\delta)$ and $\hat{\ell} = \ell_0(1-\delta)$. 
\end{minipage}
\end{table}

Opposite-sign biases produce 3.7$\times$ larger absolute error on average and near-zero coverage at high $\kappa$. Thus, removing cancellation maximizes amplification and causes coverage collapse, exactly as predicted by the cross-term structure in the bias decomposition.

\newpage

\clearpage
\sloppy
\section{Replication and Implementation Details}
\label{app:replication}

This section documents the materials and procedures required to replicate the results in the main paper and this appendix.

\paragraph*{Code Availability}
\begin{enumerate}
    \item \textbf{Repository:} All replication code is available at \url{https://github.com/gsaco/dml-diagnostic}.
    \item \textbf{Version:} To reproduce the results exactly as reported, use the repository state corresponding to the latest release tag associated with this submission.
    \item \textbf{License:} The code is licensed under the MIT License.
\end{enumerate}

\paragraph*{Software Requirements}
\begin{enumerate}
    \item \textbf{Language:} Python 3.11 or higher.
    \item \textbf{Dependencies:} Exact package versions are listed in \texttt{requirements.txt} in the repository root. 
    \item \textbf{Installation:}
\begin{verbatim}
python -m venv .venv
source .venv/bin/activate
pip install -r requirements.txt
\end{verbatim}
\end{enumerate}

\paragraph*{Data Access}
\begin{enumerate}
    \item \textbf{LaLonde Data:} The empirical application uses the LaLonde (1986) job training dataset as processed by Dehejia and Wahba (1999).
    \item \textbf{Source:} The data is downloaded automatically by the script \texttt{src/data.py} from the NBER archive. No manual download or registration is required.
\end{enumerate}

\paragraph*{Replication Instructions}
\begin{enumerate}
    \item \textbf{Single Entrypoint:} The repository includes a master script \texttt{run\_all.py} that executes all experiments and generates all outputs.
\begin{verbatim}
python run_all.py
\end{verbatim}
    \item \textbf{Output Location:} All figures and tables are written to the \texttt{results/} directory. Please refer to the Script-to-Output Mapping below for specific filenames.
    \item \textbf{Execution Time:} The full replication takes approximately 5--10 minutes on a standard laptop.
\end{enumerate}

\paragraph*{Script-to-Output Mapping}
The following scripts (located in \texttt{notebooks/}) generate the specific tables and figures reported in the text:
\begin{enumerate}
    \item \texttt{corrupted\_oracle\_analysis.py}:
    \begin{enumerate}
        \item \textbf{Main Paper:} Figures 1 and 2 (\texttt{figure1\_bias\_amplification.pdf}, \texttt{figure2\_coverage\_analysis.pdf}).
        \item \textbf{Appendix:} Tables A.1--A.6 (derived from \texttt{corrupted\_oracle\_results.csv} and \texttt{corrupted\_oracle\_aggregates.csv}).
    \end{enumerate}
    \item \texttt{lalonde\_application.py}:
    \begin{enumerate}
        \item \textbf{Main Paper:} Figure 3 (\texttt{lalonde\_forest\_plot.pdf}) and the LaLonde baseline estimates table (\texttt{lalonde\_baseline\_results.csv}).
    \end{enumerate}
\end{enumerate}

\paragraph*{Computational Details}
\begin{enumerate}
    \item \textbf{Randomness:} Seeds are fixed for deterministic splits; minor floating-point differences may occur across platforms.
    \begin{enumerate}
        \item Corrupted Oracle simulations use base seed \texttt{42}.
        \item LaLonde application uses base seed \texttt{42}.
    \end{enumerate}
    \item \textbf{Cross-Fitting:} All reported DML estimators use $K=5$ fold cross-fitting with random splits determined by the fixed seeds.
\end{enumerate}
\fussy

% Reproducibility additions: requirements.txt existing/verified; run_all.py existing/verified.

\end{document}